\documentclass{aa}
\usepackage{latexsym}
\usepackage{natbib}
\usepackage{graphicx}
\usepackage{times}  \DeclareSymbolFont{operators}{OT1}{ptm}{m}{n}
\newcounter{IonCS}
\renewcommand{\ion}[2]{\setcounter{IonCS}{#2}#1\,{\scshape{\roman{IonCS}}}}
\newcommand{\sect}[1]{Sect.\,\ref{#1}}
\newcommand{\sects}[1]{Sects.\,\ref{#1}}
\newcommand{\fig}[1]{Fig.\,\ref{#1}}
\newcommand{\figs}[1]{Figs.\,\ref{#1}}
\newcommand{\tab}[1]{Table\,\ref{#1}}

\newcommand{\NNN}[1]{{{#1}}}

\begin{document}

%=============================================================================
% TITLE
%=============================================================================
%
\title{Asymmetries of solar coronal extreme ultraviolet emission lines} 

%\titlerunning{}
%\authorrunning{}

\author{H. Peter}
%\offprints{H.~Peter}

\institute{Max Planck Institute for Solar System Research (MPS),
           37191 Katlenburg-Lindau, Germany, peter@mps.mpg.de}

%\date{}
\date{Received 15 March 2010 / Accepted 26 April 2010}

\abstract%
%
%---CONETXT---(optional)-----------------------------------------------------
{%
The profiles of emission lines formed in the corona contain information on the dynamics and the heating of the hot plasma. Only recently has data with sufficiently high spectral resolution become available for investigating the details of the profiles of emission lines formed well above $10^{6}$K. These show enhanced emission in the line wings, which has not been understood yet.
}
%---AIMS----------------------------------------------------------------------
{%
We study the underlying processes leading to asymmetric line profiles, in particular the responsible plasma flows and line broadening mechanisms in a highly filamentary and dynamic atmosphere.
}
%---METHODS-------------------------------------------------------------------
{%
Line profiles of \ion{Fe}{15} formed at 2.5\,MK acquired by the Extreme ultraviolet Imaging \NNN{Spectrometer} (EIS) onboard the Hinode solar space observatory are studied using multi Gaussian fits, with emphasis on the resulting line widths and Doppler shifts.}
%---RESULTS-------------------------------------------------------------------
{%
In the major part of the active region, the spectra are best fit by a narrow line core and a broad minor component.  The latter contributes some 10\% to 20\% to the total emission, is about a factor of 2 broader than the core, and shows strong blueshifts of up to 50\,km/s, especially in the footpoint regions of the loops. On average, the line width increases from the footpoints to the loop top for both components. A component with high upflow speeds can be found also in small restricted areas.
}
%--CONCLUIONS---(optional)----------------------------------------------------
{%
The coronal structures consist of at least two classes that are not resolved spatially but only spectroscopically and that are associated with the line core and the minor component. Because of their huge line width and strong upflows, it is proposed that the major part of the heating and the mass supply to the corona is actually located in source regions of the minor component. It might be that these are identical to type II spicules.
The siphon flows and draining loops seen in the line core component are consistent with structures found in a three-dimensional magneto-hydrodynamic (3D MHD) coronal model.
Despite the quite different appearance of the large active region corona and small network elements seen in transition region lines, both show similar line profile characteristics. This indicates that the same processes govern the heating and dynamics of the transition region and the corona.
}
%-----------------------------------------------------------------------------
%
%----------------------------------------------------------------------------

\maketitle

%---------------------------------------------------------------------------
% FIGURE Context
%---------------------------------------------------------------------------
\begin{figure*}
%
%\sidecaption
%\includegraphics[bb=56 283 566 462,clip=true]{eis_dc_context_lin}
\includegraphics[bb=56 283 566 657]{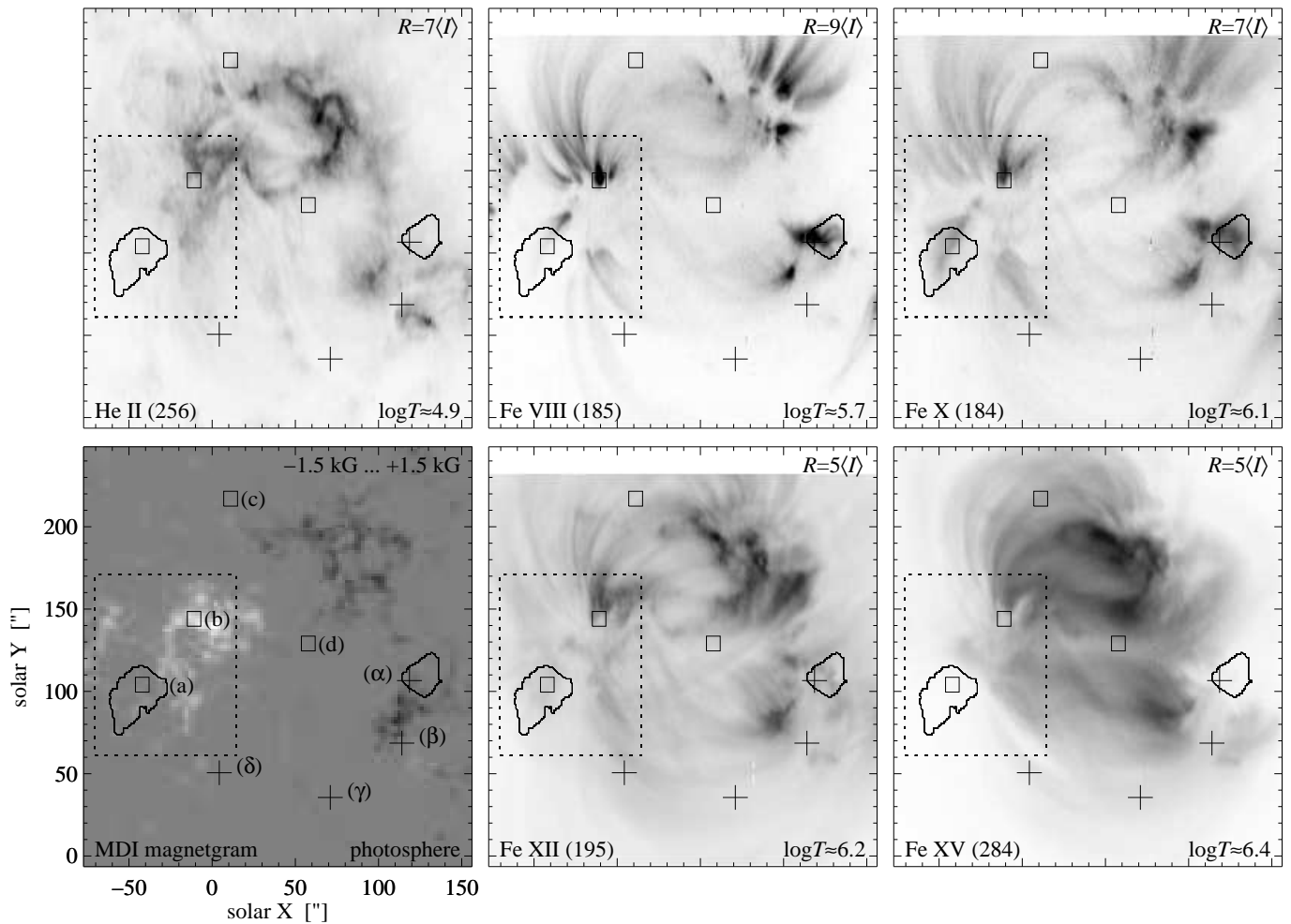}
\caption{%
Intensity maps in selected emission lines and photospheric magnetogram. The line formation temperatures (peak of ion fraction, ${\log}T$) is given in each panel. The images are plotted on an inverse linear scale, scaled with respect to the median intensity in each map, $\langle{I}\rangle$, with white representing a value $1/R$ and black $R$ where $R$ is given in the plots in units of $\langle{I}\rangle$.
The maps shown here are recorded co-spatially and co-temporally. The lower right panel shows the radiance in the \ion{Fe}{15} line that is analyzed in this paper in depth.
The lower left panel shows the MDI magnetogram of the photosphere taken at 19:15 UT during the EIS scan (linearly scaled from $-$1.5\,kG to $+$1.5\,kG.
The contour lines show two areas with high-velocity components; i.e., the minor component of a free double Gaussian fit shows a blueshift higher than 55\,km/s \NNN{when smoothed over 11$\times$11 pixels} (cf.\ \sects{S:free.double} and \ref{S:high.velo.outflow}). The rectangles show areas with average spectra plotted in \fig{F:avg.spectra}, and the crosses indicate the positions of the individual spectra (at one spatial pixel) shown in \fig{F:ind.spectra}. The dotted rectangle indicates the region used by \cite{Hara+al:2008} for their average spectrum ($\approx$85\arcsec$\times$110\arcsec). The intensities are plotted on a logarithmic scale in \fig{F:context.log} to show the patterns in the regions with low emission.
\label{F:context}
}
\end{figure*}
%---------------------------------------------------------------------------

%%%%%%%%%%%%%%%%%%%%%%%%%%%%%%%%%%%%%%%%%%%%%%%%%%%%%%%%%%%%%%%%%%%%%%%%%%%%
\section{Introduction}
%%%%%%%%%%%%%%%%%%%%%%%%%%%%%%%%%%%%%%%%%%%%%%%%%%%%%%%%%%%%%%%%%%%%%%%%%%%%

Understanding the dynamics of the plasma in the corona of the Sun is pivotal for unveiling the processes that govern the heating of the outer atmosphere of the Sun to temperatures of millions of degrees. Imaging instruments can give vital information on the evolving structures. However, they only provide information on the apparent motions, which do not necessarily have to be real plasma motions, and they are broad band in wavelength, thereby mixing several spectral lines formed at different temperatures. Spectroscopy provides details on the emission line profiles and thus supplies crucial information for investigating the thermal and dynamic structure of the coronal plasma.

Various scenarios and models make detailed predictions on the spectral profiles to be observed. Only selected aspects should be mentioned here. The nanoflare-heating models of \cite{Patsourakos+Klimchuk:2006} based on 1D hydrodynamic loop models predict high-temperature upflows of hot plasma following transient heating events (introduced adhoc in the model). These would show up as a weak additional component in the wing of emission lines formed well above 10$^6$K with a blueshift of some 100\,km/s (see their Fig.\,3). Based on their observations, \cite{DePontieu+al:2009:roots.of.heating} propose a scenario for the mass cycle of the active region corona where hot plasma related to type II spicules is propelled upwards to supply mass to the corona (which subsequently rains down after cooling). This scenario would also imply a strongly blueshifted component in the coronal emission lines. Similarly, \cite{Tu+al:2005} or, more recently, \cite{Tian+al:2010} have found evidence of the plasma in open field regions near the limb above temperatures of several 100.000\,K being accelerated to form the open wind, with blueshifts of some 30 km/s seen in lines formed at 2\,MK. If (locally) open regions on the disk, e.g. adjacent to an active region, were to produce a similar outflow, this could show up in the spectra as excess emission in the blue wing of the emission line, too.

Line profiles originating in the transition region from the chromosphere to the corona have been intensively investigated using vacuum ultraviolet spectrometers such as HRTS \citep[High Resolution Telescope and Spectrograph;][]{Brueckner+Bartoe:1983} or SUMER \citep[Solar Ultraviolet Measurements of Emitted Radiation;][]{Wilhelm+al:1995}. These provided a high spectral resolution of about 20\,000 or more allowing them not only to derive the shifts of the line centroid and width, but also to detect peculiarities in the line profiles. 
The most noticeable features in the transition region line profiles are found during explosive events \citep{Dere+al:1989:expl.events}: two distinct satellites to the blue and red indicating a bi-directional flow, most likely induced by a small reconnection event. The spatial and temporal evolution of these were first described by \cite{Innes+al:1997}, confirming the nature of a bi-directional flow.
\citet{Kjeldseth-Moe+Nicolas:1977} were the first to show that transition region spectra show enhanced emission in the line wings. \cite{Peter:2000:sec:err} presented evidence that these enhanced wings are predominantly found above the chromospheric network and are best fitted by a double Gaussian with a narrow line core and a broad second (or wing) component. A double Gaussian fit with one broad component is even more significant than a triple Gaussian fit with two narrow components accounting for the wing emission \citep{Peter+Brkovic:2003}. This questions the previous interpretation of the wing excess as caused by small counterparts of explosive events \citep{Dere+Mason:1993}. The width of the broad wing component seems to increase monotonically with line formation temperature \citep{Peter:2001:sec}, which is consistent with line broadening due to magneto-acoustic waves.

Our knowledge of the detailed spectral profiles of lines formed in the hotter corona is more limited by instrumentation in the era of SoHO \citep[Solar and Heliospheric Observatory;][]{Domingo+al:1995} and before. The Coronal Diagnostic Spectrometer  \citep[CDS;][]{Harrison+al:1995} did not provide sufficient spectral resolution to reveal details in the line profiles (except for extreme cases, e.g.\ flares), and the SUMER spectra of lines formed at hot temperatures are problematic in terms of signal-to-noise ratio.
This situation changed with the the Extreme ultraviolet Imaging Spectrometer \citep[EIS;][]{Korendyke+al:2006,Culhane+al:2007} onboard the Hinode solar space observatory \citep{Kosugi+al:2007}. While the spectral resolution of about 4000 is worse than SUMER or HRTS, the good signal-to noise ratio provides the possibility of studying the details of the line profile. 

%---------------------------------------------------------------------------
% FIGURE line shift and width
%---------------------------------------------------------------------------
\begin{figure*}
%
%\sidecaption
\includegraphics[bb=56 283 566 722]{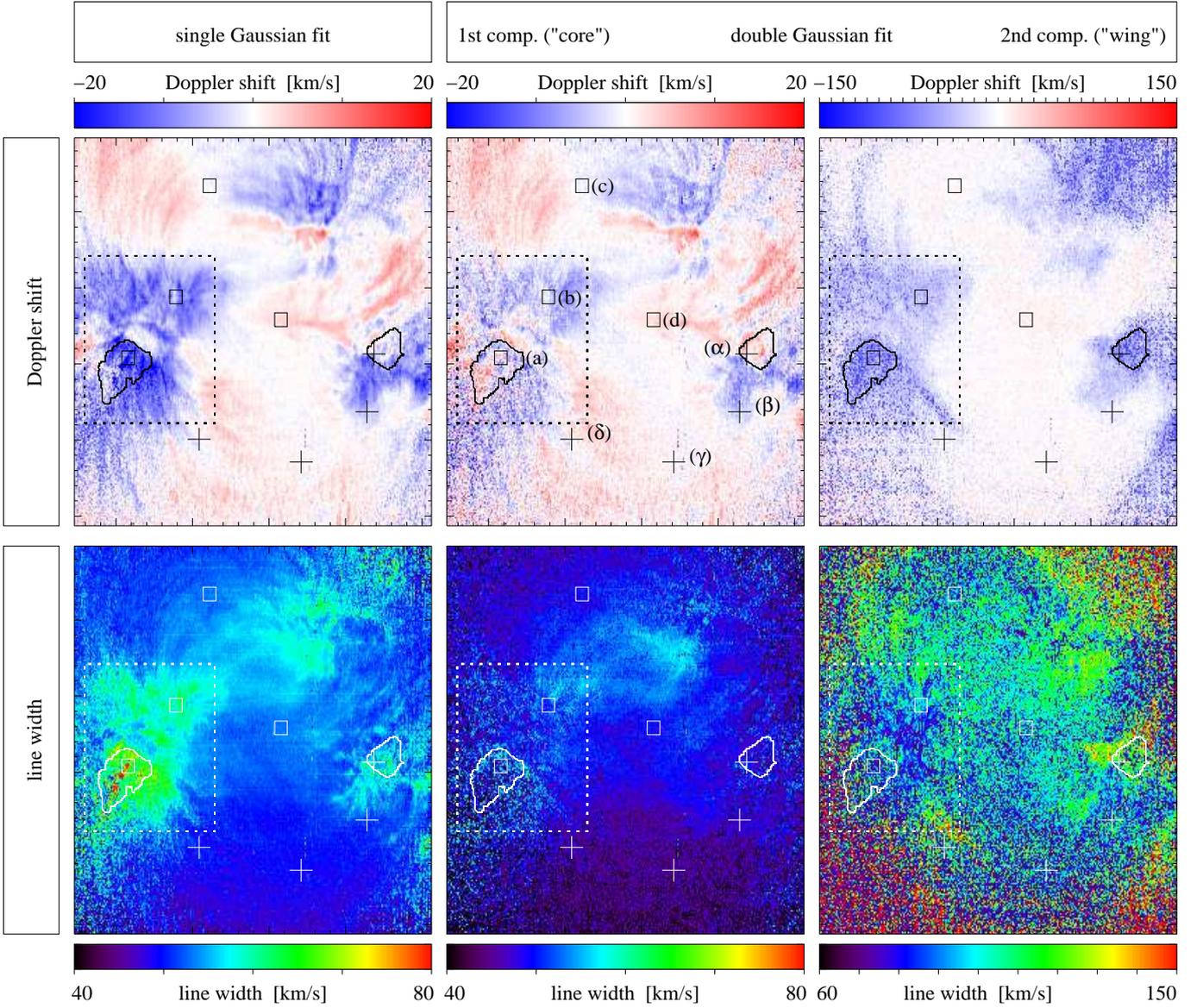}
\caption{%
Spatial maps of line shift and width for \ion{Fe}{15} (284\,{\AA}). These are shown for single Gaussian fits (left column), as well as for the core component and the minor (wing) component for a free double Gaussian fit.
The panels are co-spatial and co-temporal with respect to \fig{F:context}.
The respective color scales for the single Gaussian and the core of the double Gaussian are the same, while the scales for the minor component of the double Gaussians span far wider ranges.
The Doppler shifts have been corrected for the tilt of the spectrograph slit and the orbital motion of the spacecraft, while the line widths (exponential widths) have not been corrected for instrumental broadening.
The crosses, rectangles, and contours have the same meaning as in \fig{F:context}.
See \sects{S:single} and \ref{S:free.double}.
\label{F:shift.width.maps}
}
\end{figure*}
%---------------------------------------------------------------------------

The first study of the asymmetries of coronal lines in an active region was conducted by \citet{Hara+al:2008} using EIS data. As a first approach they performed single Gaussian fits to two emission lines and found deviations from a single Gaussian, showing excess emission in the blue wing near loop foot points. Consequently they interpret their findings by favoring the nanoflare heating model by \cite{Patsourakos+Klimchuk:2006}.  However, as demonstrated in this paper, a closer inspection of the data by a more advanced data processing shows a more detailed picture. Indeed, evidence is found for \NNN{upflows near the loop footpoints, but these are probably not consistent with the nanoflare scenario}.

The paper is organized as follows. First the data and their reduction are discussed in \sect{S:obs} before the multi Gaussian line profile fits are analyzed in \sect{S:multi.Gauss}. Subsequently, the results are discussed with respect to the high-velocity outflows in the active region periphery (\sect{S:high.velo.outflow}), to the heating and mass supply in the active region (\sect{S:broad.component}), and to the self-similarity of processes in the active region corona and the quiet Sun network transition region (\sect{S:tr}). Finally, the results are put into the context of recent three-dimensional magneto-hydrodynamic (3D MHD) coronal models in \sect{S:loop.model} and the conclusions summarized in \sect{S:conclusions}.

%%%%%%%%%%%%%%%%%%%%%%%%%%%%%%%%%%%%%%%%%%%%%%%%%%%%%%%%%%%%%%%%%%%%%%%%%%%%
\section{Observations and data processing}   \label{S:obs}
%%%%%%%%%%%%%%%%%%%%%%%%%%%%%%%%%%%%%%%%%%%%%%%%%%%%%%%%%%%%%%%%%%%%%%%%%%%%

To ensure a good comparison with existing work the present study deals with the same data set as analyzed previously by \citet{Hara+al:2008}, but now with a new more complex method to investigate the spectral profiles. A raster scan is analyzed covering about 234{\arcsec}$\times$256{\arcsec} near disk center (cf.\ \fig{F:context}) including active region NOAA 10938. The 1{\arcsec} wide slit was used with a sampling of 1\arcsec/pixel along the slit and a raster step of about 1{\arcsec} to build the map shown in \figs{F:context} and \ref{F:shift.width.maps}. The exposure time was 30\,s. The data were taken on 18 Jan.\ 2007 starting at 18:12 UT, and it took about 135 minutes to complete the raster scan.
The maps shown in \fig{F:context} were shifted along the slit ($y$-direction) to account for the spatial misalignment between the two EIS detectors (hence the white stripes at the top of some maps; Kamio 2010%
\footnote{Private communication. A publication is in preparation.}%
).

\begin{figure*}
%
%\sidecaption
\includegraphics[bb=56 283 566 679]{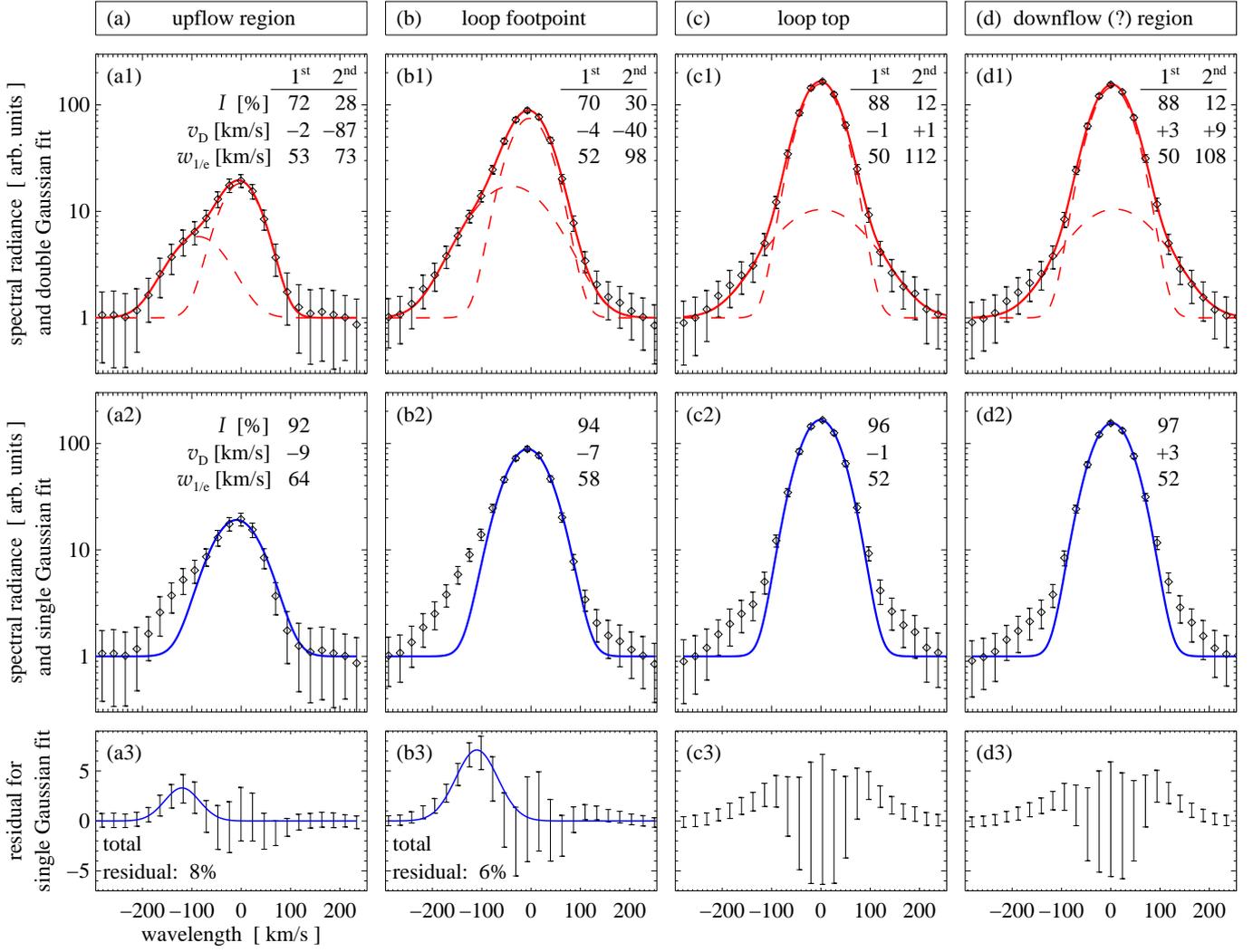}
\caption{%
Average spectral profiles of \ion{Fe}{15} (284\,{\AA}).
These are averages of the regions shown by small squares in \fig{F:context} labeled (a) -- (d), each covering 10\arcsec$\times$10\arcsec (100 individual spectra). The spectra are shown on a logarithmic scale to emphasize the excess in the line wings.
The top and middle rows show the spectra (diamonds) along with the errors (bars) normalized to the continuum intensity. (The spectra in each column are identical, e.g.\ panels a1 and a2.) The respective panels in the top row show the double Gaussian fit (solid) along with the two components (dashed), the middle column shows the single Gaussian fit (solid). The numbers with the plot give the contribution of the respective component to the total line intensity $I$, the Doppler shift $v_{\rm{D}}$ and the exponential width $w_{\rm{1/e}}$.
The lower row shows the residual of the respective single Gaussian fits. For regions (a) and (b) these residuals show a roughly Gaussian profiles with peaks near $-$120\,km/s as indicated by the solid lines (panels a3, b3; numbers give the fraction of the residual when compared to the total line intensity).
See \sects{S:single} and \ref{S:free.double}.
\label{F:avg.spectra}
}
\end{figure*}

Because the goal of this study is to investigate the line profiles originating from the hot part of the corona, this study will concentrate on the \ion{Fe}{15} line at 284\,{\AA}, formed at about 2.5\,MK under ionization equilibrium conditions. Some sample spectra are shown in \figs{F:avg.spectra} and \ref{F:ind.spectra}.
\NNN{%%%%%%
As discussed in \sect{S:blends} and appendix \ref{S:app.blends} there are potential blends in the blue wing of the \ion{Fe}{15} line by \ion{Fe}{17} and \ion{Al}{9}. However, under active region conditions, as well as for quiet Sun or flare conditions, these should contribute substantially less than the line wing excess found for \ion{Fe}{15}, and these blends are at higher blueshifts than the additional spectral component needed to fit the \ion{Fe}{15} line profile.
}%%NNN%%% 
Thus the line asymmetries discussed in this paper cannot be attributed to line blends.

The raw data were processed using the procedures and calibration data supplied in SolarSoft\footnote{http://www.lmsal.com/solarsoft.} by the instrument team. This included removing the CCD pedestal and dark current, flagging problematic pixels, as well as calibrating and estimating errors. For further details see \cite{Young+al:2009}.

For investigating the line profile, single and double Gaussian fits with a continuum have been applied. As the goal is to detect additional spectral components accounting for small excess in the line wings, a reliable fitting procedure has to be used, as normal fitting algorithms (e.g. based on hill climbing) may fail. Thus the genetic algorithm-based optimization method described by \cite{Charbonneau:1995} is used. This method also allows constraining the line fit parameters, e.g., to force the line fit to find solutions only with one spectral component at high blue shifts. The genetic algorithm is a global method, which is why it is effective in finding the \emph{global} minimum of a given function (here the difference between line fit and data). This contrasts with local methods, which are prone to getting stuck in local minima. This is of interest especially if the data to be fitted are noisy or the fitting problem is not determined very well \citep[see e.g.\ the analysis by][]{Peter:1999full,Peter:2000:sec:err}. The latter is the case here, especially when fitting a double Gaussian (seven free parameters) to a line profile with 24 wavelength pixels across the profile (cf. \figs{F:avg.spectra} and \ref{F:ind.spectra}).
To estimate the errors, a local optimization method is used (gradient expansion algorithm as provided by the curvefit procedure of IDL, Interactive Data Language). Using the results from the genetic algorithm as an initial guess, the local method can provide a good fit and automatically returns the standart deviation for the fit parameters.

%Following the genetic algorithm based optimization a standart (local) optimization method is applied using the results from the genetic algoritm as initial values for the parameters. Here the mpfit package%
%%
%\footnote{The mpfit package was written for the Interactive Data Language (IDL) and can be used to perform a Levenberg-Marquardt least-squares optimisation. It was written by Craig Markwardt, http://cow.physics.wisc.edu/$\sim$craigm/idl/fitting.html.}
%%
%is used as it allows to constrain the parameter range to values close to the genetic algorithm output. This hybrid scheme allows a faster convergence and returns error estimates for the best fit parameters (cf.\ \sect{S:error}). 

%---------------------------------------------------------------------------
% FIGURE average spectra
%---------------------------------------------------------------------------
%---------------------------------------------------------------------------

The Doppler shifts have to be corrected for the line shifts induced by the orbital motion of the spacecraft, as well as for the inclination of the slit. For this we used the Doppler shift of the line core, i.e., of the core component as determined from the double-Gaussian fits. First the slit inclination is corrected for according to the SolarSoft routines (by P.\,Young), then the orbital variation is determined by averaging the raster map along the slit ($y$-direction), by this determining the sinusoidal-like variation in time that corresponds to the variation along the $x$-direction in the raster map. This average line shift along the $x$-direction (viz.\ in time) is then subtracted from the originally determined Doppler maps. In the absence of a possibility for an absolute wavelength calibration, the average shift (of the line core) is set to zero. Thus the Doppler shifts discussed here are only relative to the average field of view. Previous studies of the quiet Sun and active regions show that the absolute line shifts at about 2\,MK can be expected to range from 5\,km/s to 10\,km/s  towards the blue \citep{Peter+Judge:1999,Teriaca+al:1999:ar}. Thus the Doppler shifts derived in this paper might have to be shifted by this amount to the blue to obtain an absolute scaling. However, most of the line shifts discussed here will be on the order of 50\,km/s, and thus this correction might not be of vital importance for the present study.

%%%%%%%%%%%%%%%%%%%%%%%%%%%%%%%%%%%%%%%%%%%%%%%%%%%%%%%%%%%%%%%%%%%%%%%%%%%%
\section{Multi-Gaussian line profile fits}      \label{S:multi.Gauss}
%%%%%%%%%%%%%%%%%%%%%%%%%%%%%%%%%%%%%%%%%%%%%%%%%%%%%%%%%%%%%%%%%%%%%%%%%%%%

When performing a multi-Gaussian fit to find an optimal representation of the observed line profile, one makes an implicit assumption, namely that a finite number of spatial regions with different plasma properties contribute to the spectrum seen in the spatial resolution element. In the case of a double Gaussian, it could be that this is ``background'' providing the main emission (the line core), and a small part of the resolution element could be filled with a strand of strong upflow (showing up in the line wing). In principle, this could also come from temporal variation, but this seems unlikely within the 30\,s exposure time data discussed here.

Other methods of investigating the line profiles share this character. If one does a single Gaussian fit and investigates the residual \citep[e.g.][]{Hara+al:2008}, one is assuming two spatial components, too. Also when comparing the emission  in the red and blue wings to characterize the line asymmetry, e.g.\ \cite{McIntosh+DePontieu:2009:upflows} assume that this asymmetry is caused by a second spatial component in the resolution element.

In contrast to this assumption that the emission is composed of distinct spatial regions alternative interpretations are also possible. The line asymmetries could be produced by opacity effects --- this is why photospheric lines are often investigated using the bi-sector technique. However, the extreme ultraviolet lines are optically thin, in general. Only very strong lines under special conditions show opacity effects, e.g., the strong \ion{C}{4} 1548\,{\AA} line near the limb \citep[e.g.][]{Mariska:1992}.
Another possibility would be that the shape of the line profile reflects the velocity distribution function of the ions, which is briefly discussed in \sect{S:velo.distribution}.
Finally an asymmetric instrumental profile could cause the asymmetric line profiles, but as discussed in \sect{S:instr.effects}, this is not probable for the data under investigation here.

In the following the properties of a single and a free double Gaussian fit to the data will be investigated first (\sect{S:single} and \ref{S:free.double}), a constrained double gaussian fit with one component forced to be at high blueshifts will be studied later (\sect{S:constrained.double}).

%===========================================================================
\subsection{A short note on the line widths}  \label{S:line.widths}
%===========================================================================

It is assumed that each component of the spectral profile can be described by a Gaussian,
\begin{equation}\label{E:Gauss}
I_{\rm{Gauss}}\;(v)  =  \hat{I} ~\exp~ \left( \frac{(v-v_{\rm{D}})^2}{w_{\rm{1/e}}^2 } \right) ~,
\end{equation}
where the wavelength is given in Doppler shift units, $v$, and the free parameters are the peak intensity, position, and width, $\hat{I}$, $v_{\rm{D}}$, and $w_{\rm{1/e}}$.
Please note that this Gaussian width $w_{\rm{1/e}}$ is related to the full width at half maximum, {\sc{Fwhm}}, of the line through
\begin{equation}\label{E:FWHM}
\mbox{\sc{Fwhm}} = 2 \sqrt{ \ln{2} } ~ w_{\rm{1/e}} ~.
\end{equation}

Following \cite{Doschek+al:2007}, the instrumental broadening of the EIS spectrometer is about 56\,{m\AA} {\sc{Fwhm}}, which translates into a Gaussian width of about 35\,km/s for the line under investigation, i.e \ion{Fe}{15} at 284\,{\AA} (cf.\ \tab{T:widths}).
The data shown in the figures in this study are not corrected for instrumental broadening, because the main aim of the paper is not a quantitative interpretation of the derived line widths. In order to characterize the line fits and to compare them with the actually observed spectral profiles, it is preferable to keep the line widths as returned from the fits. Where necessary the respective values corrected for instrumental broadening are quoted.
In principle (assuming Gaussian instrumental profiles), the ``solar'' line width is given by $(w_{\rm{fit}}^2 - w_{\rm{instr}}^2)^{1/2}$, where $w_{\rm{fit}}$ is the line width returned by the fitting procedure and $w_{\rm{instr}}$ the instrumental width.

For comparison, \tab{T:widths} also lists the thermal line width of \ion{Fe}{15} at 284\,{\AA}, assuming that the line is formed near the temperature of maximum ion fraction, along with the total line width to be expected if thermal broadening would be the only broadening mechanism.

%
%The instrumental broadening of EIS is about 56\,m{\AA} \citep[full width at half maximum, {\sc Fwhm;}][]{Doschek+al:2007}. At 284\,{\AA}, the wavelength of a strong \ion{Fe}{15} line, this corresponds to 59\,km/s {\sc Fwhm}. 
%As this dominates the thermal broadening of the line,  we can well expect to find some peculiarities in EIS spectral observations, i.e.\ deviations from a (single) Gaussian shape, in the wings of the lines about more than 30\,km/s (i.e. half of the instrumental broadening) away from line center.

%---------------------------------------------------------------------------
% TABLE line widths
%---------------------------------------------------------------------------
\begin{table}
\caption{Line widths related to \ion{Fe}{15} (284\,{\AA}).
\label{T:widths}}
\begin{tabular}{lrcc}
\hline
\hline
                   &                      &  {\sc Fwhm}  & exponential\,width
\\
                   && [km/s] & [km/s]  %\multicolumn{2}{c}{[km/s]}
\\
\hline
instrumental width$^a$ & $w_{\rm{instr}}$     &      59      &      35
\\
thermal width$^b$      & $w_{\rm{thermal}}$   &      45      &      27
\\
\multicolumn{2}{l}{
total width ~ $(w_{\rm{instr}}^2+w_{\rm{thermal}}^2)^{1/2}$}
                                          &      75      &      45
\\
\hline
\end{tabular}
\begin{tabbing}
$^a$ \= According to \cite{Doschek+al:2007}: {\sc Fwhm} $=$ 56\,m{\AA}; at 284\,{\AA}.
\\
$^b$ \> $w_{\rm{thermal}}^2=2k_{\rm{B}}T/\mu$ \= with line formation temperature ${\log}T[{\rm{K}}]{=}6.4$\\
     \>\> and molecular weight $\mu=55.8\,$amu.
\end{tabbing}
\end{table}
%---------------------------------------------------------------------------

%---------------------------------------------------------------------------
% FIGURE sample spectra
%---------------------------------------------------------------------------
\begin{figure*}
%
%\sidecaption
\includegraphics[bb=56 283 566 606]{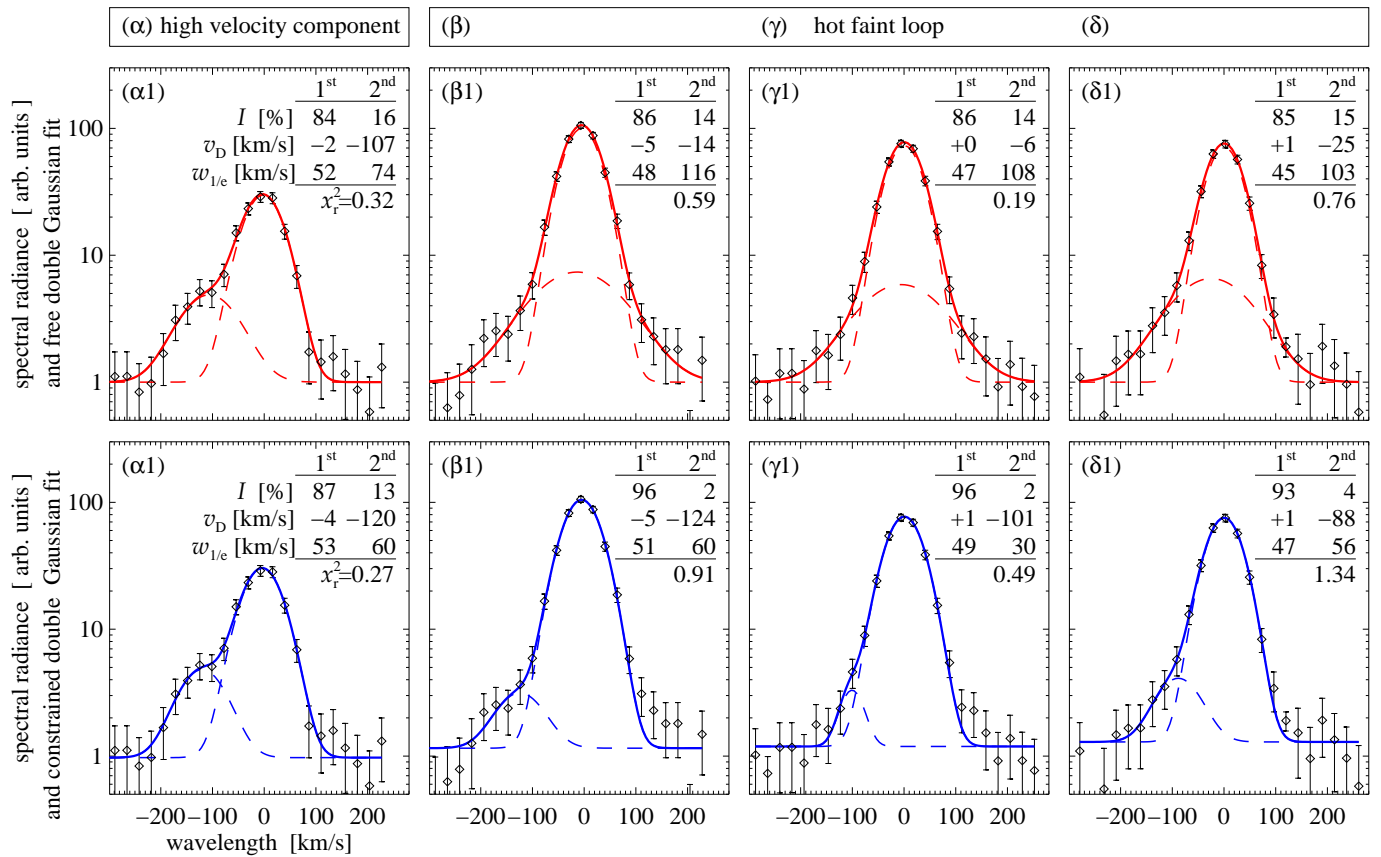}
\caption{%
Individual spectral profiles of \ion{Fe}{15} (284\,{\AA}). 
These were recorded in one spatial pixel ($\approx$1{\arcsec}$\times$1{\arcsec}) shown by crosses in \fig{F:context} labeled ($\alpha$) -- ($\delta$). As in \fig{F:avg.spectra} the profiles are shown on a logarithmic scale to emphasize the line wings.
The top row shows the line profiles along with the free double Gaussian fits (cf.\ \sect{S:free.double}), the bottom row for the constrained double Gaussian fits where the minor component was forced to have a blueshift of at least 70 km/s. (cf.\ \sect{S:constrained.double}).
The dashed lines show the two components of the fit, the solid line the sum of the two components. The diamonds show the actual spectra with the bars representing the errors. The observed spectra in each column are identical (e.g.\ the spectra in $\alpha1$ and $\alpha2$).
As in \fig{F:avg.spectra}, the numbers in the plots give the line fit parameters, now accompanied by the values for $\chi_{\rm{r}}^2$.
See \sects{S:free.double} and \ref{S:constrained.double}.
\label{F:ind.spectra}
}
\end{figure*}
%---------------------------------------------------------------------------

%===========================================================================
\subsection{Single Gaussian fits and excess emission in the line wings}  \label{S:single}
%===========================================================================

As a first step, just a single Gaussian fit is preformed. The leftmost column of \fig{F:shift.width.maps} shows the resulting maps in line shift and width. The maps are very similar to those shown in \cite{Hara+al:2008}, of course.
As already noted by \cite{Hara+al:2008}, the spectral profiles are in general not well fit by a single Gaussian and show excess emission in the line wings. In the sample (average) spectra shown by \cite{Hara+al:2008}, the line excess seems to be concentrated in the blue wing of the line, and they interpret this as the result of chromospheric evaporation as expected, e.g., following impulsive heating events.  This would be expected according to the  models of \cite{Patsourakos+Klimchuk:2006}.

The region used for the average spectrum in the \cite{Hara+al:2008} study is outlined in \figs{F:context} and \ref{F:shift.width.maps} (dotted rectangles). While \cite{Hara+al:2008} note that this area represents the footpoint region of the active region loop system, it is clear from inspection of \fig{F:context} that the situation is far more complex.
Therefore in the present paper smaller areas covering about 10\arcsec$\times$10{\arcsec} will be examined. They contain 100 spatial pixels and are outlined in \figs{F:context} and \ref{F:shift.width.maps} (small rectangles). Only region (b) is covering a loop footpoint. Area (a) is in a fan-like structure seen in the cooler lines (\ion{Fe}{8} and \ion{Fe}{10}), while region (c) is located in the middle of a hot loop (seen in \ion{Fe}{15}). Area (d) is located in the middle of the active region, where the coronal emission is only very weak. The important observation is that these small areas show quite different line profile characteristics.

%%
%% Hara et al average spectrum: x = -115 ... -30  ,  y = -60 ... +50 in relative coordinates 
%%                              (where 0,0 is the middel of the raster...
%%

In \fig{F:avg.spectra} average spectra in these selected areas (a) -- (d) are displayed. The middle row shows the raw spectra along with the single Gaussian fits and the lowest row displays the residuals of the fits. In contrast to \cite{Hara+al:2008}, the spectra are plotted on a logarithmic scale to give a better representation of the line wing excess. It is clear that only a fraction of all spectra show an excess emission in the blue wing alone.
Only the case (a) labeled ``upflow region'' in the fan-like structure shows a clear blue-wing-only excess (indicated by the Gaussian for the residual; \fig{F:avg.spectra}, panel a3). In general, excess emission is found in both wings of the line, sometimes with more weight to the blue (as in panel b3), mostly with relatively comparable excess in both wings (panels c3 and d3), but almost never with an excess in only the red wing.

%---------------------------------------------------------------------------
% FIGURE Histogram chi-squared & line contribution
%---------------------------------------------------------------------------
\begin{figure*}
\sidecaption
\includegraphics[bb=56 283 396 438]{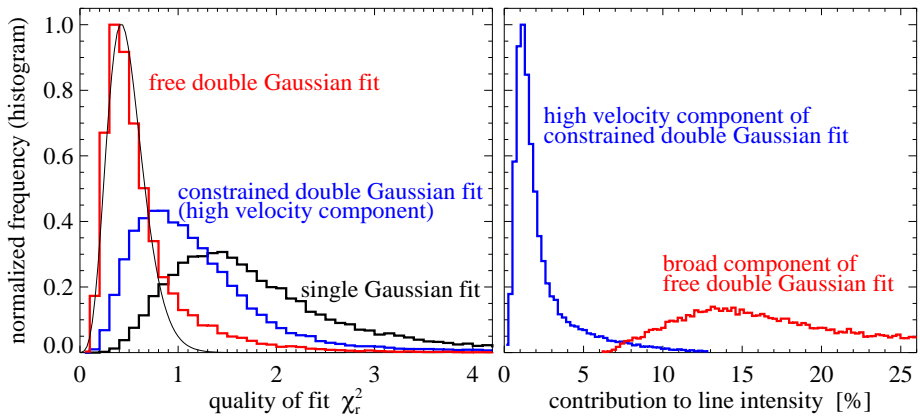}
\caption{
Distribution of fit quality and contribution to line intensity.
The left panel shows the histograms for the distribution of the quality of the fits $\chi^2_{\rm{r}}$ to the \ion{Fe}{15} line profile. The thin solid line shows the theoretical distribution (with $\chi^2_{\rm{r}}$ scaled by 0.5 to account for the probable error in the EIS error determination).
\newline
The right panel shows the contribution of the second minor components of the constrained an the free double Gaussian fits to the total line intensity.
See \sects{S:best.fit} and \ref{S:broad.component}.
\label{F:chi.histogram}
}
\end{figure*}
%---------------------------------------------------------------------------

The asymmetries of the line profile will also influence the line shifts and widths derived from the single Gaussian fits, of course. For example, there will be the tendency for a line profile with excess emission in the line wing to result in a larger line width for a single Gaussian fit: the fitting procedure will push the line width to values that are as high as possible to account for the excess line wing emission, even though it cannot fully account for the excess.

The single Gaussian fits show a systematic change in the line width from low values near the middle (top) of the hot loop system (near rectangle c in \fig{F:context}, \ion{Fe}{15} panel, and in \fig{F:shift.width.maps}, lower left panel) to the footpoints of the loops (respective rectangles b). This has been already noted by \cite{Hara+al:2008}, who interpret this as evidence of increased heating near the loop footpoints as advocated e.g.\ by \cite{Aschwanden+al:2007}.

However, this observational finding of \cite{Hara+al:2008} is only an artifact of the excess emission in the line wings. When performing a double Gaussian fit, the line core component shows the opposite trend in the line width, with smaller widths near the loop footpoints (cf. lower middle panel of \fig{F:shift.width.maps})!
This does not rule out the possibility of the heating being concentrated near the footpoints, as some mechanism has to account for the excess line wing emission, but it clearly shows that more care has to be taken when analyzing EUV emission line spectra with a spectral resolution and a signal-to-noise ratio good enough to see line asymmetries.

%===========================================================================
\subsection{Free double Gaussian fits}  \label{S:free.double}
%===========================================================================

Double Gaussian fits have been performed to quantify the contribution of the wing excess. First a free fit is described before a fit for a restricted parameter range is discussed in the following section. Actually, for the genetic algoritm used here for the fit (cf.\ \sect{S:obs}), a range of parameters has to be specified. For the free fit, the parameter range was chosen broad enough not to constrain the fit in any way.

Without constraints, the line profile is typically decomposed into a major narrow component accounting for the line core and a minor component accounting mostly for the wing excess, since it is much broader than the core component. Examples for the free double Gaussian fits can be found in \fig{F:avg.spectra} (top row) for average spectra in about 10{\arcsec}$\times$10{\arcsec} regions and in \fig{F:ind.spectra} (top row) for individual spectra recored in one spatial pixel ($\approx$1{\arcsec}$\times$1{\arcsec}).

In general, the free double Gaussian fits are much better than the single Gaussian fits discussed above, with the $\chi_{\rm{r}}^2$ on average a factor a three smaller (\fig{F:chi.histogram}, left panel). The minor component typically contributes about 10\% to 20\% to the total line emission (right panel). Inspection of the individual fits to the spectra, of which here only four are shown (\fig{F:ind.spectra}, top row), shows that the free double Gaussian fit indeed provides a good model to fit the line profile (cf.\ \sect{S:best.fit}).

The spatial maps for line shift and width for both components are shown in \fig{F:shift.width.maps} in the middle and right columns. Given that the minor component contributes less than 1/5 to the line, one would expect the line shifts of the core component for the double Gaussian fit and the shift of the single Gaussian fit to be comparable. Mostly this is correct, but not in regions where the minor component shows strong blueshifts. Here the single Gaussian fit shows strong blueshifts, while the double component fit reveals that the line core is mostly redshifted (!), and the minor component is a satellite in the blue wing (for sample spectra, see panels a1 and $\alpha$1 in \figs{F:avg.spectra} and \ref{F:ind.spectra}). This strong outflow component is restricted to two small areas and is discussed in more detail in \sect{S:high.velo.outflow}.%
\footnote{
\NNN{Some of the spectra in these areas might be affected by line bends, which is discussed in \sect{S:blends} and in more detail in appendix \ref{S:app.blends}.}
}
Again, this emphasizes the need for a proper multi-Gaussian line fit.

The line widths of both the core and the minor component are larger in the middle part of the active region than in the footpoint regions; i.e., it seems that the lines are broader in the upper parts of the loops than near their feet. This contrasts the results of the single Gaussian fit. At first sight this seems contradictory; however, this is due to the minor component generally showing stronger blueshifts in the footpoint regions. The line width and its implication for the coronal heating and dynamics are discussed in detail in \sect{S:broad.component}.

%---------------------------------------------------------------------------
% FIGURE constrained double Gaussian fit
%---------------------------------------------------------------------------
\begin{figure*}
\sidecaption
\includegraphics[bb=56 283 396 455]{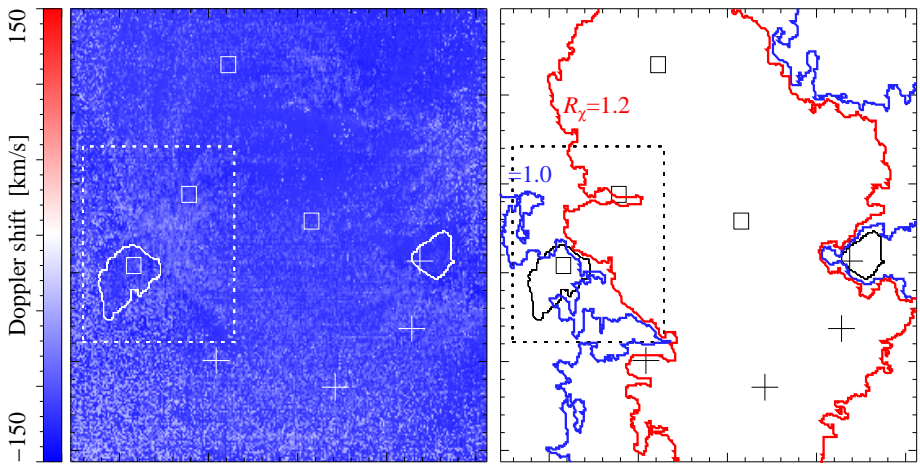}
\caption{%
Constrained double Gaussian fits and their significance.
The left panel shows the Doppler map for the minor (high-velocity) component of the Gaussian fits constrained to have a high-velocity component.
\newline
The right panel shows contour lines of the ratio $R_{\chi}$ of the $\chi^2_{\rm{r}}$ for the constrained double Gaussian fit divided by the $\chi^2_{\rm{r}}$ for the free double Gaussian fit. Same field of view as in \figs{F:context} and \ref{F:shift.width.maps}.
See \sect{S:constrained.double}.
\label{F:high.velo}
}
\end{figure*}
%---------------------------------------------------------------------------

%===========================================================================
\subsection{Forcing a high-velocity component}  \label{S:constrained.double}
%===========================================================================

The double Gaussian fits reveal a high-blueshifted component in restricted areas at the periphery of the active region, while in the central part covering most of the area the second minor component shows moderate blueshifts combined with large line widths. Considering a spectrum as shown in column b of \fig{F:avg.spectra}, one could still argue that a double Gaussian fit with a high-blueshift component would do as good a job.
Thus a double Gaussian fit was performed with two constraints: (1) The minor component has to be shifted by more than 60\,km/s to the blue and (2) its Gaussian width should be smaller than 65\,km/s (i.e.\ not larger than the maximum width of the core).

The resulting Doppler shifts of the minor high-velocity component are in the range of 100\,km/s to 140\,km/s: if the minor component is forbidden to also account for the excess in the red wing, then it moves far out to the blue. This is depicted in the left panel of \fig{F:high.velo} showing the map of shifts of the minor high-velocity component. Very high blueshifts on the order of 100\,km/s are found everywhere. That the shifts are always below 160\,km/s confirms that the line profile of \ion{Fe}{15} investigated here is not blended by the \ion{Fe}{17} line (cf.\ \sect{S:obs}). As \ion{Fe}{17} would be the strongest potential blend, it can be concluded that the present study is not affected by line blends.

When looking at individual spectra, it quickly becomes clear that the fit constrained to have a high-velocity component is not doing very well in general. It simply fails to properly account for the red wing emission (see panels $\beta$1, $\gamma$1, $\delta$1 in \fig{F:ind.spectra}). Only in those restricted regions where the free double Gaussian fit shows a strongly blueshifted component does the constrained fit do a good job (\fig{F:ind.spectra}, column $\alpha$). This is underlined by the $\chi_{\rm{r}}^2$ values given in the panels of \fig{F:ind.spectra}, as well as by the histogram for the $\chi_{\rm{r}}^2$. In general (outside the high-blueshift regions), the $\chi_{\rm{r}}^2$ for the constrained fit is about a factor of 2 larger than for the free double Gaussian fit. This is even true when looking at the footpoint regions of the hot loops, e.g.\ in region (b) outlined in \figs{F:context} and \ref{F:shift.width.maps}.

To emphasize that the major part of the active region is better fit by a free double Gaussian, contour lines for the ratio $R_{\chi}$ of the $\chi_{\rm{r}}^2$ for the constrained fit divided by the one for the free double Gaussian fit are plotted in \fig{F:high.velo}.
Inside the red contour line for $R_{\chi}{=}1.2$, the free double Gaussian fit is certainly more significant than the fit with the constrained high-blueshift component, and this region covers the whole part of the active region with bright coronal emission (cf.\ lower right panel of \fig{F:context}). Outside the $R_{\chi}{=}1.0$ contour, where the constrained fit is better than the free fit, only the high-velocity regions are found (thick black contours). These are enclosed by a type of pocket of the $R_{\chi}{=}1.0$ contour. ($R_{\chi}$ is also below 1 in areas with very low signal in the coronal line, where the fit is not reliable for both the constrained and the free Gaussian fits.)

From this one can conclude that a high-blueshift component is indeed only present in the two restricted areas outlined in \figs{F:context} and \ref{F:shift.width.maps} by the contour lines, which are located at the periphery of the active region. In the major part of the active region, the minor component of the double Gaussian is broad (more than twice the line width of the core) and shows moderate blueshifts ranging from zero to some\,50 km/s (cf.\ \fig{F:two.d.histo}).

%===========================================================================
\subsection{What is the best fit to the line profile?}
						\label{S:best.fit}
%===========================================================================

In the preceding sections, the reduced $\chi^2_{\rm{r}}$ was used to judge the significance of the fit. This is defined as \citep[e.g.][]{Bevington+Robinson:1992}
\begin{equation}\label{E:chi}
\chi^2_{\rm{r}} = \sum\limits_{i=1}^{N} ~\frac{1}{N{-}f}~
	\frac{ d_{i}^{\,2} - m(v_{i})\,^2  }{\sigma_{i}^{\,2}}  ~.
\end{equation}
Here $d_{i}$ denotes the data (i.e.\ spectral radiance) taken at spectral position $i$ and $m(v_{i})$ is the model value of the fit at the Doppler shift $v_{i}$ of that position, where $m$ is a (single or double) Gaussian as defined in (\ref{E:Gauss}) plus a continuum. The respective measurement errors are given by $\sigma_{i}$. Data at $N{=}24$ spectral positions have been used and the fit has $f{=}7$ (4) degrees of freedom for a double (single) Gaussian fit.
Assuming that the individual measurements (i.e.\ the spectra of the spatial scan) are independent (and the errors are normally distributed), in a large enough data set the values of $\chi^2_{\rm{r}}$ should be distributed with a peak around $\chi^2_{\rm{r}}{\approx}1$ \citep[for basics see e.g.][]{Bevington+Robinson:1992}.

The $\chi^2_{\rm{r}}$ distribution for the best fit, i.e.\ the free double Gaussian, shows a peak near $\chi^2_{\rm{r}}{\approx}0.5$ and roughly follows the theoretical distribution \citep[][Sect.\ 11.1]{Bevington+Robinson:1992} if scaled by a factor of 2 (cf.\ left panel of \fig{F:chi.histogram}).
It is unclear why exactly the peak of the $\chi^2_{\rm{r}}$ distribution is only 0.5, i.e. less than 1. One reason could be that the errors $\sigma_{i}$ of the data are overestimated. The errors returned by the EIS data calibration software are basically Poisson errors, but they also account for other effects such as CCD read-out noise \citep[see][]{Young+al:2009}. If these errors were overestimated by some 40\% (${\approx}1{-}\sqrt{2}$, this would account for the factor of 2 in the $\chi^2_{\rm{r}}$ distribution (cf.\ Eq.\ \ref{E:chi}). A detailed analysis of the measurement errors of EIS would be needed to fully investigate this, which is beyond the scope of this paper. 
The assumption of a normal distribution of the errors for the $\chi^2$ test is a potential problem, too. It remains open as to what extent this will influence the distribution of $\chi^2_{\rm{r}}$ and in which direction. Possible problems in the determination of the measurement error and for the underlying assumptions for the $\chi^2$ test show that it is no surprise that the peak of the $\chi^2_{\rm{r}}$ distribution is not at 1.

%Another possibility would be that the the assumptions for the $\chi^2$-test are not properly fulfilled, i.e.\ that the data values are not normally distributed. As the histogram of the peak intensity of the emission line is not a Gaussian (cf. \fig{F:int.histo}), this certainly violates the assumptions for the $\chi^2$-test, but it remains open to what extent this will influence the distribution of $\chi^2$ and in which direction. Possible problems in the determination of the measurement error and for the underlaying assumptions for the $\chi^2$ test show that it is no surprise that the peak of the $\chi^2$ distribution is not at 1.

Despite these problems, one can compare the $\chi^2_{\rm{r}}$ distributions for the different fits. From \fig{F:chi.histogram} (left panel) it is clear, that the free double Gaussian performs best, while the single Gaussian fit clearly fails to be a good representation of the data since it cannot account for the wing excess. The double Gaussian fits constrained to have a high-blueshift component perform better than the single Gaussian fits, but because the high-velocity component is only present in (two) restricted areas, the constrained Gaussian fit falls short compared to the fit with two free Gaussians.

%===========================================================================
\subsection{Errors for best-fit parameters}
						\label{S:error}
%===========================================================================

When using the genetic algorithm-based optimization one sould have to investigate the curvature of the $\chi^2_{\rm{r}}$ surface in the multidimensional parameter space to determine the errors of the line-fit parameters. As this is a complex problem of its own, a local optimization method is used with the results from the genetic algorithm as the initial guess to estimate the uncertainties of the fit (cf.\ \sect{S:obs}). This should provide a good approximation for the errors of the fit.

In the case of the free double Gaussian fit, the mean errors for the line shift are ${\pm}5$\,km/s for the core and ${\pm}7$\,km/s for the minor component. The respective values for the errors of the Gaussian width are ${\pm}7$\,km/s and ${\pm}16$\,km/s.
It is clear that the errors are greater for the minor component, since the shift and the width of a weak (mostly broad) component are more difficult to determine.
Please note that these are just the errors for the line profile fit so they do not include other error sources.

%%%%%%%%%%%%%%%%%%%%%%%%%%%%%%%%%%%%%%%%%%%%%%%%%%%%%%%%%%%%%%%%%%%%%%%%%%%%
\section{Discussion}
%%%%%%%%%%%%%%%%%%%%%%%%%%%%%%%%%%%%%%%%%%%%%%%%%%%%%%%%%%%%%%%%%%%%%%%%%%%%

%===========================================================================
\subsection{Possible effects of line blends}
						\label{S:blends}
%===========================================================================

%---------------------------------------------------------------------------
% TABLE: blends
%---------------------------------------------------------------------------
\begin{table}
\NNN{%%%%%%%
\caption{Contribution of possible blends to \ion{Fe}{15} at 284\,{\AA}.\label{T:blends}}
\begin{tabular}{l@{~~}r@{}l@{~~}rcrrrr}
\hline\hline
                   &   rest        &&  relative  
                                   &&\multicolumn{4}{c}{modeled contribution  [\%]\,$^a$}\\
\cline{6-9}
\multicolumn{2}{r@{}}{wavelength}  &&    shift$^b$  &&       & active &  quiet  & coronal \\
line                 & $[$\AA$]$   &&   $[$km/s$]$  && flare & region &    Sun  &    hole \\
\hline
\ion{Fe}{17}         & 284.010 &$^c$   & $-$158     &&   2.8 &    0.1 & $<$0.1  &     0.0 \\
\ion{Al}{9}          & 284.015 &$^d$   & $-$152     &&   0.2 &    0.6 &    2.1  &    83.7 \\
\ion{Fe}{15}         & 284.160 &$^e$   &      0     &&  97.0 &   99.3 &   97.8  &    16.3 \\
\hline
%\multicolumn{8}{l}{According NIST data base wavelength of \ion{Al}{9} is 285.015\,\AA}\\
%
\end{tabular}
\\[0.4ex]
$^a$~~Contribution modeled using the Chianti atomic data base.
\\
$^b$~~Doppler shift according to difference in rest wavelength. % to \ion{Fe}{15}.
\\
$^c$~~From NIST atomic spectra data base \citep{Ralchenko:2008}.
\\
$^d$~~From NIST (as $^c$); based on laboratory measurements by \cite{Valero+Goorvitch:1972}, who estimate uncertainty to be  $\pm0.04$\,{\AA} ($\pm$40\,km/s).
\\
$^e$~~Based on solar observations as provided by \cite{Brown+al:2008}, uncertainty is given as $\pm0.004$\,{\AA} ($\pm$4\,km/s). Within errors consistent with NIST data base (observed and theoretical calculated values).
}%%%NNN%%%%%
\end{table}
%---------------------------------------------------------------------------

\NNN{%--------------------------------------------------------------------------

According to the Chianti atomic data base \citep[v6.0;][]{Dere+al:2009}, there are two potential blends in the range of 300\,km/s to the red or blue of the \ion{Fe}{15} line investigated here: an \ion{Fe}{17} line formed in hot plasmas such as found in flares \citep{Feldman+al:1985,DelZanna+Ishikawa:2009} and an \ion{Al}{9} line formed at somewhat lower temperatures than \ion{Fe}{15} \citep[e.g.,][]{Brown+al:2008}.
To estimate the potential influence of these blends the relative contribution of the three lines was calculated using Chianti for different situations (using different differential emission measure curves). The results are summarized in \tab{T:blends}.
The \ion{Fe}{17} line can be ruled out since it would play a role only in a flare, which is not considered here.
Under active region conditions, the \ion{Al}{9} line should contribute less than a percent, and even in the quiet Sun, it should be a factor of 50 weaker than \ion{Fe}{15}. 
As discussed in \sect{S:free.double}, about 10\% to 20\% of the line intensity is excess emission in the line wings (cf. \fig{F:chi.histogram}, right panel for free double Gaussian), and the excess emission shows shifts well below the approx.\ 150\,km/s shift of the potential blend (cf.\ \fig{F:avg.spectra}, \fig{F:shift.width.histo}, and \tab{T:blends}).

Thus it can be concluded that in the parts of the active region which are bright in \ion{Fe}{15} the blending by \ion{Al}{9} is negligible. Any deviations from a single Gaussian line profile which are larger than about 5\% cannot be attributed to blends, but have to result from flows along the line of sight. 
However, in the areas with low emission in \ion{Fe}{15}, which might be relatively cool, the blending by \ion{Al}{9} could be present. Going to the extreme case of a coronal hole with maximum temperatures below 1\,MK, the \ion{Al}{9} emission would even dominate \ion{Fe}{15} (cf.\ \tab{T:blends}).
In the regions where the minor component of the free double Gaussian fit shows high blueshifts at low \ion{Fe}{15} emission, in about 1/3 of the area the fit might pick up the blend.
This is discussed in detail in appendix \ref{S:app.blends}.
The issue of a potential \ion{Al}{9} blend affects only the discussion in \sect{S:high.velo.outflow}, and then only partly, as the blend is only found in a fraction of the region discussed there.

}%NNN------------------------------------------------------------------------------

%===========================================================================
\subsection{Instrumental effects causing line asymmetries?}
						\label{S:instr.effects}
%===========================================================================

If an asymmetry of the instrumental profile caused the excess in the line wings, this effect should be roughly the same across the field of view. However, the broad minor components have quite different strengths, shifts, and widths. For example, in the average spectra shown in panel b1, c1, and d1 in \fig{F:avg.spectra}, the relative shift between the core and the minor component spans a range from 7\,km/s to 35\,km/s. This is considerable compared to the width of the line core and on the same order as the (exponential) width of the instrumental profile (assuming it were Gaussian). This wide range of relative shifts between minor and core components is a clear indication that the effect is non-instrumental, especially as the different relative shifts are associated with solar structures, i.e., loop footpoints and loop tops (cf.\ panels b1 and c1 in \fig{F:avg.spectra}).

%===========================================================================
\subsection{Asymmetric velocity distribution function of the ions?}
						\label{S:velo.distribution}
%===========================================================================

An asymmetric velocity distribution function of the emitting ions could be the reason for the line asymmetries, too. Usually it is assumed that the high densities in an active region ensure strong enough collisional coupling to have Maxwellian distribution functions. However, deviations from this seem to be possible, as shown e.g.\ by \cite{Bourouaine+al:2008.non.resonant} for non-resonant ion heating by Alfv\'en waves in the low corona. The extended wings of the ion distribution in velocity space would directly translate into the observable emission line profile. Were this the cause of the asymmetric line profile, the interpretation would be radically different from the interpretation discussed here. Instead of spatially separated source regions for the line core and the minor component (cf.\ \fig{F:cartoon}), the emission could originate in a monolithic structure. The asymmetry would not arise from structures in real space, but from peculiarities in velocity space. Future investigations will have to show whether this process is relevant in active regions, too, and if it indeed is compatible with the observations presented here.

%===========================================================================
\subsection{High-velocity upflows in the active region periphery}
						\label{S:high.velo.outflow}
%===========================================================================

The minor components of the double Gaussian fits fall into two categories. (1) In the high-emission part of the field of view, i.e., in the core of the active region, the minor component is broad and shows moderate shifts up to some 50 km/s, which is discussed in \sects{S:broad.component} and \ref{S:tr}. (2) Distinct from this are two small regions at the periphery of the active region with narrower minor components at high blueshifts of \NNN{about 40\,km/s to 60\,km/s}, which is discussed in this section. 
\NNN{%
There is also a fraction of area (about 1/3) in these regions that shows shifts of some 100\,km/s, but it is not clear that this is truly a component of \ion{Fe}{15} or a blend (see discussion in appendix \ref{S:app.blends} and histogram in \fig{F:FeX.histo}).
}
%
% following statement is wrong. Deleted.
% These high-velocity spectra show up as a distinct peak in the two-dimensional histogram of % width vs.\ shift (lower left part of \fig{F:two.d.histo}b).

Both high-velocity regions are located in areas with low coronal emission at 2.5\,MK (${\log}T{=}6.4$) \NNN{and are outlined by contour lines near (a) and ($\alpha$) in \figs{F:context} and \ref{F:shift.width.maps}}. It is important to note that the line core emission, accounting for about 80\% to 90\% of the total line emission, shows very low blueshifts, or even redshifts in these high-velocity areas (cf.\ \fig{F:shift.width.maps}, middle and right panels in top row). In both cases one might consider the line core component being emitted in the major part of the structure, where plasma is sitting in a more or less static fashion.
As seen in \fig{F:context}, both areas are outside the photospheric magnetic field concentrations, and they do not seem to be related to the emission from the chromosphere and transition region seen in \ion{He}{2}. However, they are quite different in terms of emission from the cooler parts of the corona.

\subsubsection{Solar wind outflow from filamentary funnels?}  \label{S:wind}

After comparison with the \ion{Fe}{8} line formed at 0.5\,MK (${\log}T{=}5.7$) in \fig{F:context}, the high-velocity region near (a) seems to be above a fan-like cool structure. This might be interpreted as a coronal funnel, the cool emission indicating the location of its footpoint, but further analysis of magnetic data would be needed for a conclusion on this.
\NNN{%
Even though not emerging from a full sunspot, this structure might also be related to a sunspot plume \citep{Maltby+al:1999} because the emission from \ion{Fe}{10} shows a redshift associated with it (cf.\ \fig{F:FeX.Doppler}).
}% 

Since the line core emission of \ion{Fe}{15} from the major part of the structure is not shifted (or only little), the high-velocity component could then be considered as ``fingers'' reaching into the main structure. These fingers could either supply mass to the structure or they could be the base of strands through which material is channeled into the upper corona or even the wind. 
%
%This latter interpretation is of particular interest when considering recent work of \tobedone{} Tian et al. (2009) on the origin of solar wind streams at the edges of active regions. 
%
Especially the high-velocity region near (a) might be a good candidate to supply mass to the wind at high velocities already close to the surface. The temperature in solar wind streams should be rather low (${<}1$\,MK) as shown for fast streams from polar coronal holes by \cite{Wilhelm+al:1998corona}, because most energy goes into acceleration instead of heating. Thus the very low \ion{Fe}{15} intensity in this high-velocity region near (a) as seen in \fig{F:context} or \ref{F:context.log} is in line with the interpretation that this represents the origin of a solar wind stream.

In contrast to previous scenarios, here the wind outflow would come from small strands that are embedded in an almost static coronal funnel. This would change the picture of a more or less homogeneous coronal funnel \citep{Gabriel:1976} at the base of the wind to a scenario where the funnel would be a filamentary structure with fast outflows and more static structures next to each other. In a way one could look at this as a smaller version of plume and inter-plume regions in (polar) coronal holes where the origin of the fast wind is found.

It should be noted that the findings presented here are quite different from the studies of \cite{Doschek+al:2008} or \cite{He+al:2010} who also find evidence of a wind outflow in the periphery of an active region. However, they only investigate single Gaussian fits. The regions with high-velocity components in the double Gaussian fits such as are found here coincide with the locations of strong blueshifts of the single Gaussian fits (cf. \fig{F:shift.width.maps}) --- the single Gaussians show the strong blueshifts because the optimization procedure does its best to fit the line excess in the blue wing and thus leads to a strong blueshift of the single Gaussian. Thus one can speculate that the \cite{Doschek+al:2008} and the \cite{He+al:2010} data actually show the same effect of a high-velocity component, which will have to be checked by re-examining their data.

\subsubsection{Nanoflare heated loops?}   \label{S:nanoflares}

The other high-velocity region located near ($\alpha$) in \fig{F:context} is located above strong emission in \ion{Fe}{8}, as well as in \ion{Fe}{10} formed at 1.2\,MK (${\log}T{=}6.1$). One might speculate wheather this high-velocity region is above a footpoint of a faint hot loop following the crosses ($\alpha$) -- ($\delta$). It is barely visible in \ion{Fe}{15} when plotted on a linear scale (\fig{F:context}), but clearer when displayed on a logarithmic scale (\fig{F:context.log}). The strong emission at lower temperatures might hint at an increased level of heating at the base of the loop, e.g.\ in the form of field line braiding that would lead to high energy deposition low in the atmosphere \citep{Gudiksen+Nordlund:2002}.

For an increased activity near the loop foot, one can expect that the braiding and twisting of the field lines will result in frequent nanoflare events in the coronal part of the loop. The observed high-blueshift component near the loop footpoint would be in line with the model prediction by \cite{Patsourakos+Klimchuk:2006} based on a model for a nanoflare heated loop. In this model a high-velocity upflow at the footpoint (i.e.\ small chromospheric evaporation) is driven by the nanoflare-induced heat input in the coronal part of the loop. \NNN{While the} relative strength of the spectral component due to the high-velocity upflow in their model is similar to what is found in the observations presented here, \NNN{the Doppler shift in the model is about twice as high.} As for the filamentary funnels, here one would expect that the upflow would happen in a (small) part of the individual strands forming the hot coronal loop, which again would be in line with the \cite{Patsourakos+Klimchuk:2006} model.

Thus one could conclude that these observations \NNN{support the} nanoflare heated coronal loop scenario as proposed by \cite{Patsourakos+Klimchuk:2006} \NNN{(assuming that the high-velocity component indeed is not due to a blend)}. However, there are also problems. These high-blueshift regions are seen in a small part of the field of view and only at the periphery of the active region, which would challenge the uniqueness of the \cite{Patsourakos+Klimchuk:2006} model. Furthermore, the loop with the high-upflow leg is quite dim, while the heart of the active region with much stronger emission from hot plasma does not show the pronounced high-velocity components. Thus it might be questionable if the process described by \cite{Patsourakos+Klimchuk:2006} were the dominant heating mechanism.

However, it should be emphasized that, in principle, the high-velocity components as predicted by \cite{Patsourakos+Klimchuk:2006} could still be hidden below the intense radiation from the main part of the active region. More precisely, it could be that the high-velocity component is not visible because other effects causing the excess in the line wings are stronger as discussed in the next section. 
Furthermore, their model is a pure (M)HD description of flows along the loop. In a more realistic model, it might be that they will find a stronger broadening of the emission from the upflows. For example, some MHD turbulence induced by the evaporation process could lead to a line broadening and might decelerate the upflow, which could be the reason for the observed properties of the broad minor component in the bight active region loops. 

Thus the present observations cannot be used to rule out their attractive proposal.
Future work will have to show whether the upflows induced by the nanoflares could be identical to the source region of the broad minor components in the active regions loops.

%---------------------------------------------------------------------------
% FIGURE shift & width histogram
%---------------------------------------------------------------------------
\begin{figure*}
\sidecaption
\includegraphics[bb=56 283 396 438]{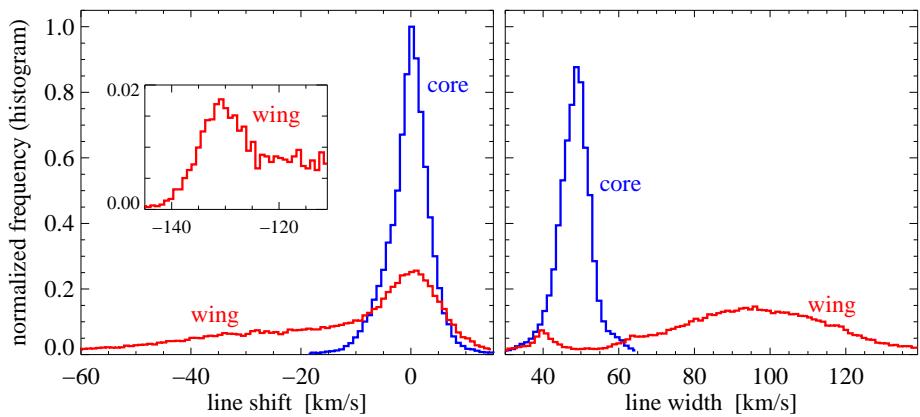}
\caption{%
Distribution of line shift and line width for free double Gaussian fits to \ion{Fe}{15} (284\,{\AA}). The distributions are shown for the line core component as well as for the minor component labeled ``wing''. The minor component has a contribution at high blueshifts, as can be seen in the inlet in the left panel, \NNN{which is most probably due to a blend (cf.\ appendix \ref{S:app.blends})}. The small distinct contribution to the line widths at very small widths (40\,km/s) is an artifact of bad fits. The line widths are not corrected for instrumental broadening. See also \fig{F:two.d.histo} and \sect{S:broad.component}.
\label{F:shift.width.histo}
}
\end{figure*}
%---------------------------------------------------------------------------

%===========================================================================
\subsection{Heating and mass supply in the active region}
						\label{S:broad.component}
%===========================================================================

\NNN{For the remainder of the paper}, the discussion turns to the major part of the active region \NNN{bright in \ion{Fe}{15}}, where most of the coronal emission originates. \NNN{This region is not susceptible to effects of line blending (cf.\ \sect{S:blends} and appendix \ref{S:app.blends}).} Here the spectra are composed of a narrow line core and a minor component. Typically, the latter contributes 10\% to 20\% to the line intensity, is a factor of two borader than the core, and shows line shifts from 0 to some 50\,km/s to the blue, while the core component shows only little shifts lower than ${\pm}10$\,km/s (cf. \figs{F:chi.histogram} and \ref{F:shift.width.histo}).

\subsubsection{Waves and velocity anisotropies}  \label{S:wave.asymm}

Both the core and the minor component show a trend toward weaker line broadening near the footpoints of the coronal structures (see \sect{S:free.double}).  This could either the stronger broadening effect higher up in the coronal loops at lower densities, or the broadening could be due to an anisotropic velocity distribution of the ions.

Assuming a constant (or only slowly varying) energy flux density, a compressible disturbance (e.g.\ magnetoacoustic wave) moving upwards along a loop would steepen in amplitude as the density decreases. This is well known in the chromosphere, where acoustic waves steepen into shocks and form the \ion{Ca}{2} grains \citep[e.g.][]{Carlsson+Stein:1997}. A similar effect could in principle be expected for the corona. In the framework of coronal seismology, models and observations (mostly from imaging instruments) are used to investigate these periodic disturbances in the corona that have recently also been found in complex 3D configurations \citep[e.g.][]{Ofman:2009}.

Longitudinal waves such as discussed recently by, e.g, \cite{DeMoortel:2009} or \cite{DeMoortel+Bradshaw:2008} do not seem to be important here, because they should produce a stronger line broadening near the loop footpoints, where the angle between the magnetic field and the observer is relatively small. Higher up, where the line of sight is perpendicular to the loop magnetic field, the broadening effect of a longitudinal wave should be very small. Thus the predicted change of line broadening would be opposite to what is observed here.

In consequence, transverse magnetoacoustic oscillations \citep[e.g., review of][]{Ruderman+Erdelyi:2009} would predict just the observed trend from loop foot to apex. However, in the models these waves are damped to heat the coronal plasma. Thus their amplitude should decrease with height in the loop, which would result in a rather small line broadening. Further modeling efforts would be needed to investigate what the models predict in terms of line broadening along a loop-like structure, which would only imply small extensions of the modeling efforts of e.g.\ \cite{Terradas+al:2008} or \cite{VanDoorsselaere+al:2009}.

Alfv\'en waves in the corona have been directly detected through coronagraph observations of the magnetic field only recently \citep{Tomczyk+al:2007}. This was the first proof that Alfv\'eninc disturbances propagate upward and downward along loop structures. As these are transverse waves, they would nicely explain why the broadening is the smallest at the footpoints where the line of sight is (more or less) parallel to the magnetic field and thus perpendicular to the velocity disturbances.
For transition region lines, \cite{McIntosh+al:2008} present a new interpretation for the observed line widths that relies on such Alfv\'enic motions superposed by longitudinal motions along the line of sight.
Alfv\'enic motions have been found for the long-known type I spicules \citep{DePontieu+al:2007.sci}, as well as for the more recently identified type II spicules \citep{DePontieu+al:2007.two.spicules}.
However, the coronal Alfv\'en waves as identified by \cite{Tomczyk+al:2007} have periods of some 5 minutes and thus would not cause a line broadening of the coronal spectra investigated here, which have 30\,s exposure time. (For a multi-stranded loop such low-frequency waves might have an effect even for shorter exposure times if many strands are in one resolution element of the detector.) It remains to be seen whether Alfv\'enic disturbances with high frequencies and high enough power exist in the corona to explain the observed line broadening.

Kinetic models for heating the corona and accelerating the wind by high-frequency Alfv\'en waves through ion-cyclotron resonances predict high temperatures of the ions, in particular, the temperature perpendicular to the magnetic field would be considerably higher than those parallel to the magnetic field, as the velocity distribution functions will be anisotropic. This has been discussed in detail for the open corona extensively e.g.\ by \cite{Marsch+Tu:1997} or \cite{Cranmer:2000}, and evidence for the direct observation of the broadening through this process close to the Sun above the limb was presented by \cite{Peter+Vocks:2003}. More recently, models for resonant and non-resonant wave-particle interaction showed that also  in closed large-scale magnetic structures strong temperature anisotropies can be present for the heavy ions \citep{Bourouaine+al:2008.non.resonant,Bourouaine+al:2008.loop}. The anisotropic velocity distribution in this kind of models with larger (average) velocities perpendicular to the magnetic field would be consistent with the line broadening being larger towards the loop top. 

As for the Alfv\'en waves discussed above, further modeling would be needed to investigate if line broadening predicted by this anisotropy would agree also quantitatively with the observations, and if the kinetic models are indeed applicable in the high-density environment of an active region as investigated here.

\subsubsection{Mass supply to coronal structures}  \label{S:mass supply}

A central question, which has not been addressed yet, concerns the nature of the line core and the minor component. Recently, \cite{DePontieu+al:2009:roots.of.heating} and \cite{McIntosh+DePontieu:2009:upflows} have looked at the asymmetry of line profiles by comparing the emission from the blue and the red wings at the same distance from the line centroid (integrating over some wavelength range). Through this they find and confirm the asymmetry as found by \cite{Hara+al:2008} in other data sets and adopt the interpretation of \cite{Hara+al:2008} that the line asymmetry is caused by a high-velocity upflow in the blue wing. They associate these strong upflows with type II spicules.
Their interpretation would be consistent with the results shown here, with some important alterations and extensions. These are mainly based on the more detailed data analysis, which allows also quantitative measures for line shifts and widths.

The upflow speeds of 50\,km/s to 100\,km/s at the loop footpoints that are derived by \cite{DePontieu+al:2009:roots.of.heating} are significantly higher than the blueshifts presented in this paper for the footpoint regions. This difference can be reconciled by inspection of column (b) of \fig{F:avg.spectra} showing an average spectrum of a 10{\arcsec}${\times}$10{\arcsec} footpoint region. In that example the proper double Gaussian fit returns a minor component with a blueshift of about 50\,km/s. In contrast first fitting a single Gaussian and then investigating the residual (panel b3) would indicate an upflow with more than 100\,km/s (with a very small line width). This difference is simply because the single Gaussian does its best to also account for the wing excess and thus overestimates the contribution from the line core. Already assuming a high-velocity component will give high-velocity narrow minor components, which are less significant than the double Gaussians with a free (broad) minor component (cf.\ \sect{S:constrained.double}). Comparing the typical Doppler shifts of the minor component near the footpoints of some 30\,km/s to 50\,km/s one can conclude that \cite{DePontieu+al:2009:roots.of.heating} might overestimate the upflow speed by a factor of 2.

More importantly, the present study can give information on the line width of the minor component, and thus provides a way to investigate the physical mechanism driving the mass upflow as discussed above in \sect{S:wave.asymm}.

Future studies comparing and combining different methods to characterize the line asymmetries have to be conducted. The method used by \cite{DePontieu+al:2009:roots.of.heating} and \cite{McIntosh+DePontieu:2009:upflows} is quick and very robust, especially if the data become noisy. The study presented here requires clean spectral profiles to give reliable results but returns more quantitative information.

%---------------------------------------------------------------------------
% FIGURE 2D histogram
%---------------------------------------------------------------------------
\begin{figure}
%
%\sidecaption
\includegraphics[bb=56 283 305 515]{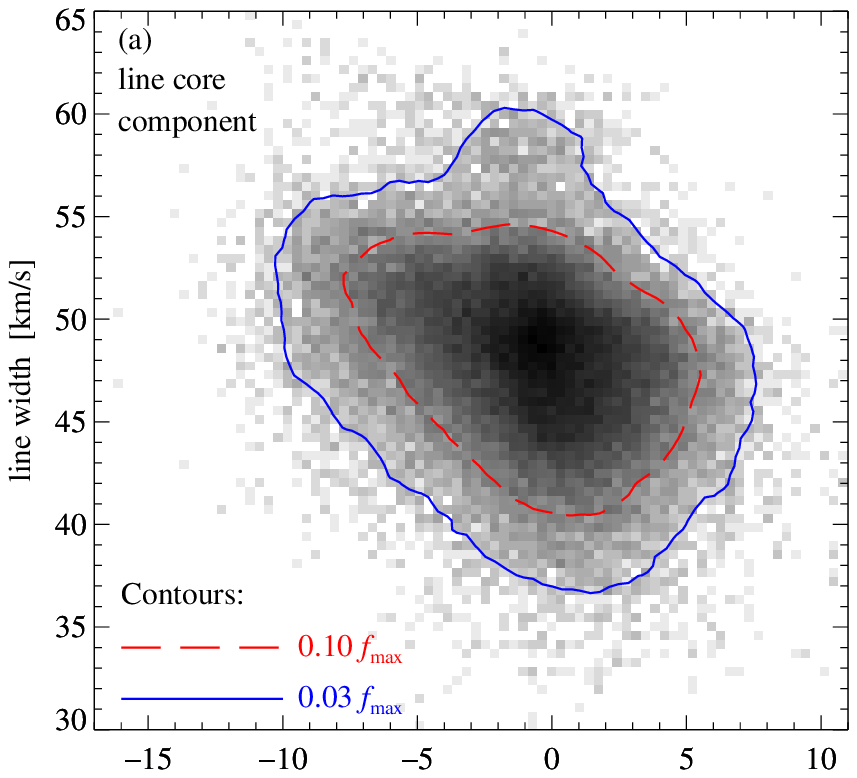}
\includegraphics[bb=56 283 305 515]{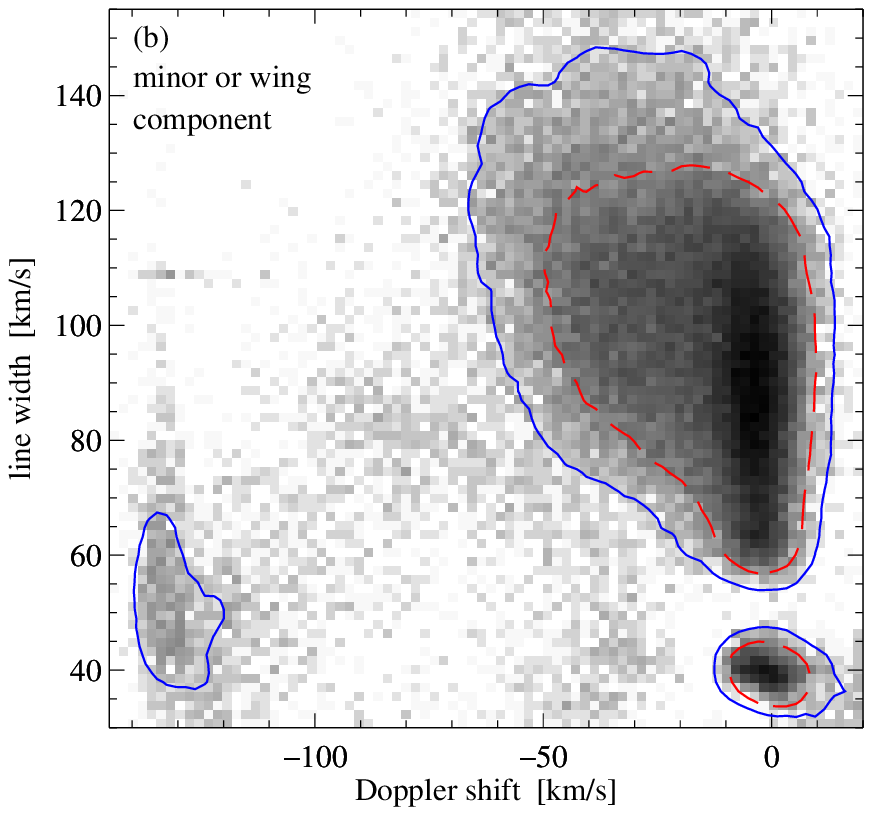}
\caption{%
Two-dimensional histograms of line width vs.\ shift for free double Gaussian fit to \ion{Fe}{15} (284\,{\AA}). The top panel (a) shows the line core component, the bottom panel (b) the minor or wing component.
The histograms are plotted on a logarithmic scale, the contours showing levels of 0.1 (dashed) and 0.03 (solid) of the maximum value.
For the minor component a contribution at high shifts with small widths \NNN{most probably is due to a blend (cf.\ appendix \ref{S:app.blends})}.
The peak with zero shift at very small width (lower right part of panel b) is an artifact due to bad fits.
See \sects{S:stat} and \ref{S:tr}.
\label{F:two.d.histo}
}
\end{figure}
%---------------------------------------------------------------------------

\subsubsection{Statistical connection between line shift and broadening} \label{S:stat}

The relation of line shifts and widths contains valuable information on the heating and dynamics of the plasma and might reveal details of the governing processes. For example a clear relation of increased line width with high Doppler shift could indicate a heating process driving a flow.

When plotting a two-dimensional histogram of line shifts and widths for the line core of the free double Gaussian fit, no clear relation can be found (\fig{F:two.d.histo}a).%
\footnote{It should be noted that for a single Gaussian fit a clear relation is found (with higher blueshifts at larger widths; not shown in this paper). This is only because the line wing excess is not properly accounted for, and thus is an artifact of the fit. It is related to the increased line widths and shifts seen for the single Gaussians in the footpoint areas (cf.\ end of \sect{S:single}).}
One might argue that there is a weak relation, and a closer inspection shows that the median width of regions with positive shifts (upflows) is ${\approx}$49\,km/s, which is a little larger than the median width of ${\approx}$47\,km/s in redshifted (downflow) areas. 
This weak relation of shift to width indicates that there is no clear connection between heating of the plasma and the associated flows in the regions where the line core of the emission profile is formed.

The situation is different for the minor component in the major part of the active region. Besides a large part of spectra with very low shifts (lower ${\pm}$10\,km/s), there is a tendency for the minor component to have a higher blueshift for increasing line widths (see \fig{F:two.d.histo}b, solid contour line). This indicates that a stronger heating in the source region of the minor component leads to increased line broadening and drives an upward flow into the corona.
This interpretation suggests that the actual mass supply to the corona is located in the source region of the minor component. This interpretation would be consistent with the proposal by \cite{DePontieu+al:2009:roots.of.heating} and \cite{McIntosh+DePontieu:2009:upflows} if the type II spicules they propose as the mass suppliers were associated with the broad minor spectral component of the coronal lines. As these type II spicules host strong Alfv\'en waves \citep{DePontieu+al:2007.two.spicules}, this seems reasonable.

%---------------------------------------------------------------------------
% FIGURE sample from 3D model
%---------------------------------------------------------------------------
\begin{figure*}
%
%\sidecaption
\includegraphics[bb=56 283 566 419]{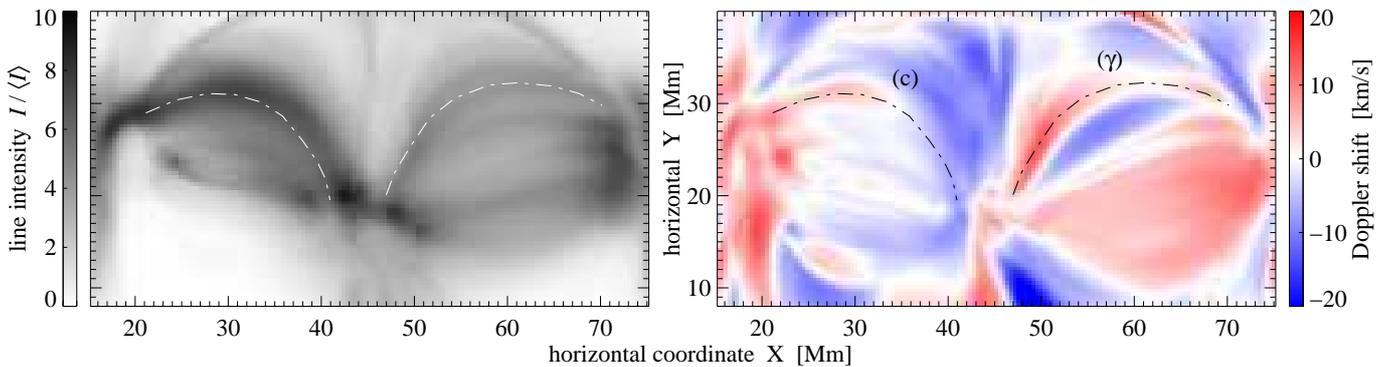}
\caption{%
Comparison to results from 3D MHD coronal model.
Displayed are the synthesized coronal emission and Doppler shifts \citep{Peter+al:2004,Peter+al:2006} for the \ion{Mg}{10} line at 625\,{\AA}. This snapshot of the time-dependent simulation shows the computational box being viewed from straight above. The intensity is normalized to its median value ${\langle}I{\rangle}$ and plotted on an inverse scale. The Doppler map is plotted on the same scale as the map of the observed line core shifts in \fig{F:shift.width.maps}.
The dashed lines indicate the location of loops showing a siphon flow (c) or a downflow on both sides ($\gamma$). These loops might be compared to the loop systems seen in the observations with their apex close to (c) and ($\gamma$) in \figs{F:context} (or \ref{F:context.log}) and \ref{F:shift.width.maps}. See \sect{S:loop.model}
\label{F:model}
}
\end{figure*}
%---------------------------------------------------------------------------

%===========================================================================
%           Relation to results for the transition region
\subsection{Self-similarity from the transition region to the corona}
						\label{S:tr}
%===========================================================================

An active region as investigated in this paper and a quiet Sun network patch at transition region temperatures are quite different in terms of spatial structure, size, and energetics. Nonetheless, numerous similarities of the spectral profile properties (in a statistical sense) can be identified. This hints at common processes on very different scales.

For transition region lines \cite{Peter:2000:sec:err,Peter:2001:sec} analyzed the spectral profiles in a similar way to the present paper, i.e., double Gaussian fits were performed. In network regions, these also typically show a narrow line core and a broad minor component. The minor component contributes some 10\% to 20\% to the total intensity and is about a factor of 2 to 3 broader than the core component, consistent with the coronal line investigated in this paper.

For the \ion{C}{4} line at 1548\,{\AA} formed at around 0.1\,MK, the relation of width to shift for both spectral components has been investigated by \cite{Peter:2000:sec:err}. The absolute values of shift and width are lower, but their relation is remarkably similar to the coronal \ion{Fe}{15} line: while no relation is found for the line core, the line shift is closely related to the line width, with low shifts for small widths.

Finally, the line width for the minor component found for the coronal line in this study falls into the trend found with the transition region lines. \cite{Peter:2001:sec} finds a monotonic increase in the non-thermal line widths for the minor component of the transition region lines (his Fig.\,7) following a power law $T^{1/4}$ (with the line formation temperature $T$). Extrapolating to ${\log}T{=}6.4$ where \ion{Fe}{15} is formed, this gives about a 100\,km/s non-thermal speed. At the footpoint regions of the active region investigated here with blueshifts of the minor component of 40\,km/s to 50\,km/s, the exponential width of the minor component of \ion{Fe}{15} is found to be about 110\,km/s to 120\,km/s (cf.\ \fig{F:two.d.histo}b). Using the instrumental and thermal widths as listed in \tab{T:widths}, the non-thermal speed of \ion{Fe}{15} is in the range of 100\,km/s to 110\,km/s, just matching the extrapolation from the transition region.

There is one major difference, though. While the medium value for the line shift of the minor components in the transition region is low (and more or less consistent with zero shift, considering the error bars), the minor components of the coronal line show a clear blueshift of up to 50 km/s (which is about 20\% of the sound speed at ${\log}T{=}6.4$). This could indicate that the actual mass feeding to the corona is located at temperatures above several $10^5$\,K, as suggested by \cite{Tu+al:2005}.

The above similarities of the statistical properties and the match of the (coronal) line width to the extrapolation from the transition region suggests that there is a common process governing the physics leading to the minor components of the line profiles all the way from temperatures of $10^4$\,K to several $10^6$\,K. In a self-similar fashion, network patches in the transition region and large active regions seem to operate in a unique way. This is remarkable because the magnetic flux in a single network patch is many orders of magnitude smaller than in an active region. Future work on more coronal lines formed at various temperatures and on different active regions will be needed to solidify this picture.

%===========================================================================
\subsection{Flows in loops matching 3D MHD model}
						\label{S:loop.model}
%===========================================================================

The observations presented here should also be discussed briefly in the context of 3D MHD models of the coronal structure and dynamics. This discussion is to show the potential of combining these observations and models. Further investigations will be needed to fully explore this.

The emission from such a 3D MHD coronal model has been synthesized by \cite{Peter+al:2004,Peter+al:2006}. This is based on heating the corona through the dissipation of currents induced by braiding the (coronal) magnetic field through photospheric convective motions \citep{Gudiksen+Nordlund:2002,Gudiksen+Nordlund:2005a,Gudiksen+Nordlund:2005b}. As these 3D coronal models cannot (yet) resolve the small structures that might host the source region of the minor components (e.g.\ type II spicules), the discussion concentrates on a brief comparison between the shifts seen in the line core component to those derived from the MHD model. In \fig{F:model} the line intensity and shift synthesized for the \ion{Mg}{10} line are displayed with the computational box viewed from above. The data are taken from \cite{Peter+al:2006}, who also provides movies of the temporal evolution.

In the observation, there are two distinct flow patterns to be found within the same active region. The hot loop system seen in the \ion{Fe}{15} line in \fig{F:context} with the apex close to (c) shows a nice siphon flow (see middle top panel in \fig{F:shift.width.maps}). This flow is probably driven by asymmetric heating at the loop feet, and a further investigation would be needed to see whether this is really supported by the line width observations. A hot faint loop can be seen in \ion{Fe}{15} with its apex close to $\gamma$ (more clearly seen with logarithmic scaling in \fig{F:context.log}). This shows downflows on both sides of the loop, indicating that the loop drains, probably because it is cooling down \citep[cf.\ different loop clases identified by][]{Ugarte-Urra+al:2009}.

Such loop structures can be modeled in the framework of 1D loop models with a prescribed heating function. For example, recently \cite{Warren+al:2010} has presented a class of 1D multi-thread models in which they successfully reproduce observations by adjusting the energy release time (and by this prescribing the heat input). In contrast to this, the temporal and spatial distributions of the energy input, as well as the resulting spatial structures, are calculated self-consistently in the 3D MHD model. (For example, at the cost of lower spatial resolution, which is also affecting the thermal conduction; the 1D and 3D models are complementary.)

In the 3D MHD models, the same flow patterns can be found as are seen in the active region studied here. In the synthetic spectral maps of \cite{Peter+al:2006}, one can find siphon flows, as well as draining loops. Examples roughly matching the observed cases are highlighted in \fig{F:model} and labeled (c) and ($\gamma$) accordingly. While these structures are present in the model at the time of the snapshot shown in \fig{F:model}, the situation is quite different only 10 minutes later. This would be consistent with the observations by \cite{Ugarte-Urra+al:2009}.

This comparison of a 3D MHD model snapshot to observations shows that the 3D MHD models can \emph{in principle} reproduce the flow patterns seen in real active regions. However, the magnetic configuration at the lower boundary of the 3D MHD model and the photospheric magnetic field for the coronal observations discussed here are different. For proof, new simulations will thus be needed. In these, the evolving magnetic field as observed in the photosphere will be used as a (time-dependent) lower boundary condition. The resulting synthetic spectra from the 3D model can then be compared directly to the actually observed coronal emission. The recently launched space-based Solar Dynamics Observatory (SDO) will provide an ideal platform for such investigations closely tying observations to 3D MHD coronal models.

%---------------------------------------------------------------------------
% FIGURE Cartoon
%---------------------------------------------------------------------------
\begin{figure*}
\includegraphics{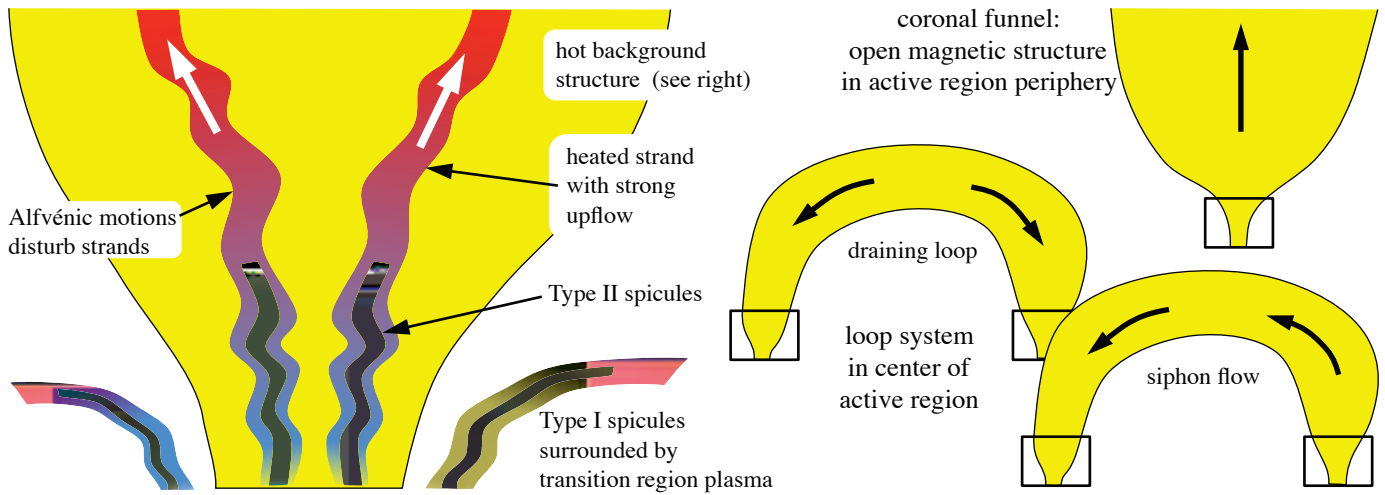}
\caption{%
Cartoon of footpoint region of coronal structure. The left panels shows cool type II spicules (as seen in \ion{Ca}{2}) embedded in the hot background plasma. These are surrounded by a thin transition region (blue) and are the basis of hot plasma flows into the corona (red). The strong heating and the Alfv\'enic motions associated with these structures lead to strongly broadened emission. This is then seen in the EUV spectrum as the broad minor component with strong blueshifts.
The heated strands associated with the type II spicules feed the mass to the coronal structure. 
The smaller images to the right depict the coronal structures possibly emerging from the footpoint region (rectangles) shown to the left.
The background structure near the feet can show upflows or downflows (depicted by the black arrows).
See \sect{S:conclusions}.
% In this study these three types can be identified.
%
\label{F:cartoon}
}
\end{figure*}
%---------------------------------------------------------------------------

%\newpage

%%%%%%%%%%%%%%%%%%%%%%%%%%%%%%%%%%%%%%%%%%%%%%%%%%%%%%%%%%%%%%%%%%%%%%%%%%%%
\section{Conclusions}					\label{S:conclusions}
%%%%%%%%%%%%%%%%%%%%%%%%%%%%%%%%%%%%%%%%%%%%%%%%%%%%%%%%%%%%%%%%%%%%%%%%%%%%

This study emphasizes the importance of information hidden in details of the spectral line profiles, which show severe deviations from a single Gaussian in coronal lines formed in active regions. In this manuscript the interpretation is adopted that (two) different spatial structures contribute to the line profile and, through this, lead to a double Gaussian profile. The line fits show that the line profile is in general composed of a narrow Gaussian accounting for the line core emission and a minor broad Gaussian component for the excess emission in the line wings.
The results for the line shift for the line core can differ significantly from a single Gaussian fit. In some areas even opposite results can be found when comparing the Doppler shifts for the line core and the single Gaussian (as in the region with fast upflow components, cf.\ \fig{F:shift.width.maps} and \sect{S:single}).

In two small areas at the periphery of the sunspot some evidence of a high upflow velocity component in the line profiles is found. This is located in regions of weak coronal emission, supposedly along low-density structures, which might be interpreted as the footpoint regions of the (slow) solar wind (\sect{S:wind}). Evidence is found that these structures are highly filamentary with small fingers of high-velocity upflows feeding the wind sticking up into the background funnel (cf.\ \fig{F:cartoon}).

In the active region, basically all spectral profiles show a minor broad component that is about two times broader than the line core component and shows systematic blueshifts up to 50\,km/s, while the line core remains more or less unshifted (cf.\ \fig{F:shift.width.histo}). The data indicate that the line core \emph{and} the minor component are narrower near the loop footpoints than at the apex, indicating some asymmetry in the broadening mechanism. This could be caused by (Alfv\'en) waves or a heating mechanism that leads to preferential perpendicular heating of the ions (\sect{S:wave.asymm}).

Comparing the present observations with previous work leads to a picture where individual strands of a loop are feeding the loop with heated material, leaving the signature of a broad blueshifted component in the line profile. After a while, these strands will no longer be heated, will slowly cool down and become part of the background loop that produces the narrow more or less unshifted emission of the line core. Now other strands are heated and fill the loop, and so forth. These heated strands might be identified with the type II spicules proposed by \cite{DePontieu+al:2009:roots.of.heating} and \cite{McIntosh+DePontieu:2009:upflows}. This is depicted in \fig{F:cartoon}. In contrast to previous studies, this analysis also provides information on the heating and dynamics of the EUV counterparts of the type II spicules, which might be the source region of the minor components, because in the present study information on line widths and shifts of the minor component is extracted (\sect{S:mass supply}).

Depending on the magnetic structure and the distribution of the heat input between the footpoints, the background loops might host siphon flows or they might be draining in response to a lower heating rate (cf.\ \fig{F:cartoon}). Both cases can be found in the present data set and have also been found in 3D MHD simulations (\sect{S:loop.model}).

No clear evidence could be found to support the nano-flare model of \cite{Patsourakos+Klimchuk:2006}. This model predicts that there is a narrow high-velocity upflow component in the blue wing (at some 100\,km/s, which is not present in the footpoint regions of the active region loops. However, further (more realistic) modeling efforts would be needed for any conclusive statement (\sect{S:nanoflares}).

The relation of line width to line shift for the line core, as well as for the minor component, is very similar in the coronal line investigated here and the transition region emission lines originating in bright network patches (\sects{S:stat}  and \ref{S:tr}).
This shows that the basic processes driving the heating and dynamics on smaller scales in the network and on larger scales in the active region are similar in nature. (At least they have to produce the same spectral signatures.)  This is supported by the match of the minor component line width of the coronal line from the active region to the extrapolation from the transition region lines from the network patch.

In summary, these observations reveal the complex flow systems in the active region corona leading to a mass cycle with small structures (probably type II spicules) feeding mass into the corona, which could host different types of flows as depicted in \fig{F:cartoon}. It has to be noted, however, that this is only one possible interpretation. Other appealing ideas, such as an asymmetric ion velocity distribution function  causing asymmetric line profiles (\sect{S:velo.distribution}), will have to be followed and checked more closely against the new observations of coronal EUV emission line profiles.

New studies that test the ideas proposed here further will have to include the analysis of both emission lines with different mass-to-charge ratios, as well as of emission lines covering a range of line formation temperatures from the upper transition region to the hot corona.

{
\acknowledgements
Sincere thanks are due to Suguru Kamio and Luca Teriaca for discussions of the data reduction and for comments on the manuscript. I would like to thank Robert Cameron for his advice concerning the $\chi^2$ distribution. I very much appreciate the comments from an anonymous referee that led to an improved discussion of the possible line blends.
Hinode is a Japanese mission developed, launched, and operated by ISAS/JAXA, in partnership with NAOJ, NASA, and STFC (UK). Additional operational support is provided by ESA and NSC (Norway).
}

%=============================================================================
% REFERENCES
%=============================================================================

%\newpage

%\def\an        {Astron. Nachr.}%
%\def\apj       {ApJ}%
%\def\apjl      {ApJ}%
%\def\apjs      {Astrophys. J. Suppl.}%
%\def\apss      {Astrophys. Space Sci.}%
%\def\jgr       {J. Geophys. Res.}%
%\def\nat       {Nature}%
%\def\solphys   {Solar Phys.}%
\def\philtrans {Phil.\ Trans.\ Roy.\ Soc.\ Lond.}%
\def\sci       {Science}

%=============================================================================
%=============================================================================
%=============================================================================
\clearpage
\Online
\appendix

\section{Blending of the \ion{Fe}{15} line}    \label{S:app.blends}

The discussion in \sect{S:blends} made clear that, for the dataset analyzed here, the \ion{Fe}{17} line will not blend the \ion{Fe}{15} line of interest, as would only be the case under flare conditions. In contrast, the \ion{Al}{9} line could potentially blend \ion{Fe}{15} in regions where the plasma is cool enough. In a coronal hole, where temperatures stay below 1\,MK the \ion{Al}{9} could even be the dominant emitter (cf.\ \tab{T:blends}). The present data set certainly does not include a coronal hole. However, in the regions where the minor component shows high velocities (contours close to (a) and ($\alpha$) in \figs{F:context} and \fig{F:shift.width.maps}), the temperatures will be significantly lower than in the main part of the active region. Nonetheless, the temperatures there can be expected to be well above 1\,MK, simply because there is still significant \ion{Fe}{12} emission (cf.\ \fig{F:context.log}), which is formed around 1.5\,MK under equilibrium conditions. According to \tab{T:blends} one might estimate the strength of this \ion{Al}{9} blend to be at the level of a couple of percent.
In the following the discussion concentrates on these low \ion{Fe}{15} intensity high-velocity minor component regions, where the biggest effect of the \ion{Al}{9} blend can be expected.

\begin{figure}
%
%\sidecaption
\includegraphics[bb=56 283 305 481]{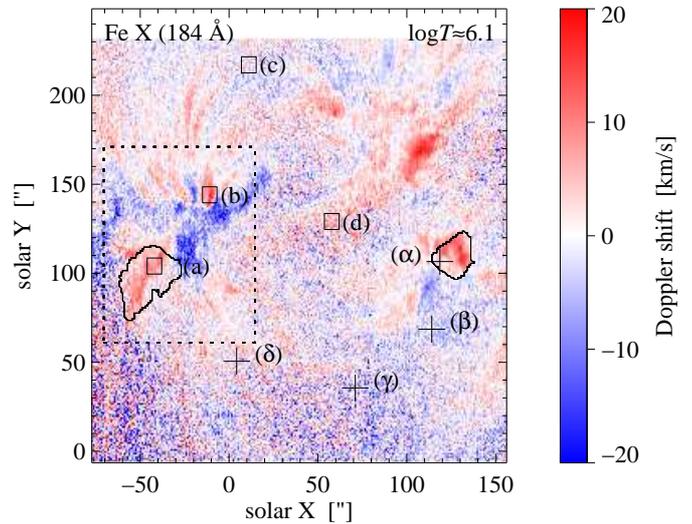}
\caption{%
Doppler shift map of \ion{Fe}{10} at 184\,{\AA}, co-spatial and co-temporal with the other maps shown in this paper. The two contour lines close to (a) and ($\alpha$) indicate the areas with a high-velocity component of the \ion{Fe}{15} line, just as in the other maps in this paper. See \fig{F:context} for further details on the labeling. The intensity map for this line is shown in the top right panel of \fig{F:context} and \fig{F:context.log}.
\label{F:FeX.Doppler}
}
\end{figure}

A major question is at what wavelength the \ion{Al}{9} blend can be expected. Based on the rest wavelengths, \ion{Al}{9} should be shifted by some 150\,km/s to the blue relative to \ion{Fe}{15} (cf.\ \tab{T:blends}). Because \ion{Al}{9} could show a net line shift, the \ion{Fe}{10} line is used as a proxy to estimate the \ion{Al}{9} net Doppler shift. Both lines form at roughly the same temperature, so they can be expected to have similar Doppler shifts. The \ion{Fe}{10} line has been recorded co-temporally and co-spatially with the \ion{Fe}{15} line, and its line radiance is shown in \figs{F:context} and \ref{F:context.log}. As the line profile of \ion{Fe}{10} is not very clear, a single Gaussian fit was performed for the line core alone. The resulting Doppler map is shown in \fig{F:FeX.Doppler}. Here, as for \ion{Fe}{15} in the main part of this paper, the wavelength is calibrated by assuming an average zero shift.

\begin{figure*}
\sidecaption
\includegraphics[bb=56 283 396 438]{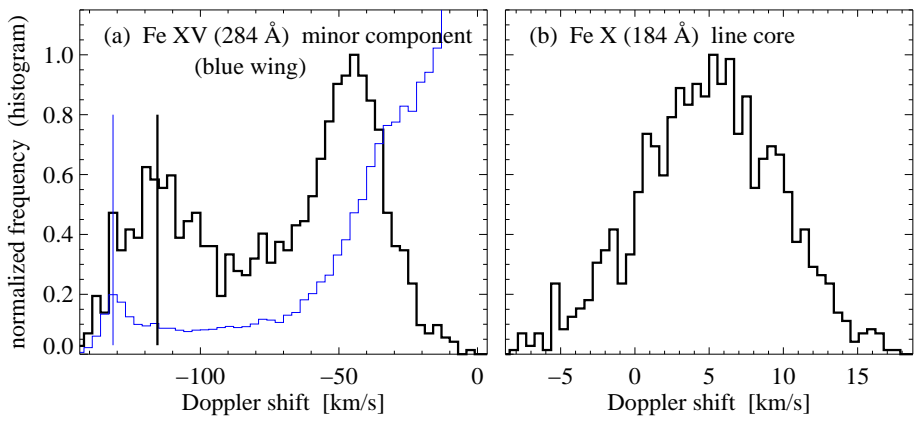}
\caption{%
Histograms of Doppler shifts in the high-velocity regions outlined by the contour lines in the maps; close to (a) and ($\alpha$) in \figs{F:context}, \ref{F:shift.width.maps}, \ref{F:FeX.Doppler}, \ref{F:context.log}.
The left panel (a) shows the histogram for the minor mostly high-velocity component of the \ion{Fe}{15} line (cf.\ top right panel in \fig{F:shift.width.maps}), while the right panel (b) shows the shift of the core of the \ion{Fe}{10} line (cf.\ \fig{F:FeX.Doppler}).
For comparison, the thin blue line in the left panel (a) shows the distribution of the \ion{Fe}{15} minor component shift in the whole field of view (same as shown in \fig{F:shift.width.histo} but scaled to show the contribution at high velocities).
The vertical lines in panel (a) indicate the location of the maxima at high velocities near  115\,km/s and 130\,km/s of the two histograms.
\label{F:FeX.histo}
}
\end{figure*}

In the region of interest (outlined by the contour lines also in \fig{F:FeX.Doppler}), the average shift of \ion{Fe}{10} is some 5\,km/s to the red, as can be seen from the histogram in \fig{F:FeX.histo} (right panel b). The core of \ion{Fe}{15} only shows a very weak shift in the region of interest, consistent with zero shift. Studies investigating the relative Doppler shift between \ion{Fe}{10} and \ion{Fe}{15} show that the \ion{Fe}{10} line is shifted some 5 to 10\,km/s to the red compared to \ion{Fe}{15}, depending on the location in the active region \citep{Tripathi+al:2009}. After combining these effects, the relative Doppler shift of \ion{Fe}{10} compared to \ion{Fe}{15} should be 0 to 5\,km/s to the red (10\,km/s maximum to give a margin). This should be comparable for \ion{Al}{9}.

Thus if assuming a maximum relative Doppler shift of 10\,km/s combined with the 150\,km/s shift of the rest wavelengths of \ion{Al}{9} and \ion{Fe}{15} mentioned above, one can estimate that the actually observed blending \ion{Al}{9} line should appear at least some 140\,km/s to the blue from the \ion{Fe}{15} line core.

%---------------------------------------------------------------------------
% FIGURE Fe X Doppler shifts
%---------------------------------------------------------------------------
%---------------------------------------------------------------------------

The histogram of the \ion{Fe}{15} minor component shifts in the high-velocity regions is shown in the left panel (a) of \fig{F:FeX.histo}. Here a bimodal distribution is found with maxima around 50\,km/s and 115\,km/s. Certainly the major part with the maximum near 50\,km/s covering about 2/3 of the area is not affected by the \ion{Al}{9} blend (even when considering the uncertainties of the \ion{Al}{9} rest wavelength of 40\,km/s as given in \tab{T:blends}).
It might be disputable if the part (1/3) at $\approx$115\,km/s is due to the blend or not. Clearly the histogram of the \emph{whole} field of view shown in \fig{F:shift.width.histo} (left panel, inlet) and by a thin blue line in \fig{F:FeX.histo} shows a clear peak near 130\,km/s, indicating that this very minor fraction of all spectra is indeed caused by the \ion{Al}{9} blend.
In the high-velocity region shown in \fig{F:FeX.histo}a (thick black), one might argue that a low side maximum is visible near 130\,km, and thus the spectra with the minor component near 115\,km/s might not be due to the blend. However, the current data do not seem to allow a final conclusion to be drawn on this.

In two different active regions, \cite{Brown+al:2008} find a component in the \ion{Fe}{15} spectra 114\,km/s and 88\,km/s to the blue of the \ion{Fe}{15} rest wavelength that contribute 3\% and 7\% to the total line profile emission (their Table 2). Given the uncertainties of the rest wavelength, it could well be that these are the same as the part of the histogram in \fig{F:FeX.histo}a close to 115\,km/s. However, it seems indisputable that the high-velocity components found in this study for the high-velocity region centering on 50\,km/s (being the majority) are \emph{not} caused by a blend.

It should be stressed here that, based on these estimations, the minor component of \ion{Fe}{15} in the bulk part of the active region that is bright in \ion{Fe}{15}, cannot come from the \ion{Al}{9} line. The minor component has shifts that are much too low (below 60\,km/s blue) and has line widths that are too large to be due to a blend. The finding that the free double Gaussian with a minor (broad) component is a better fit to the line profile than a constrained Gaussian with a high-velocity component (\sect{S:constrained.double}) provides further evidence that in general an \ion{Al}{9} blend is not a concern in the bright part of the active region.

%---------------------------------------------------------------------------
% FIGURE Fe X histograms
%---------------------------------------------------------------------------
%---------------------------------------------------------------------------

%\clearpage

\setcounter{section}{2}
\setcounter{figure}{0}

%---------------------------------------------------------------------------
% FIGURE Context
%---------------------------------------------------------------------------
\begin{figure*}
%
%\sidecaption
\includegraphics[bb=56 283 566 657]{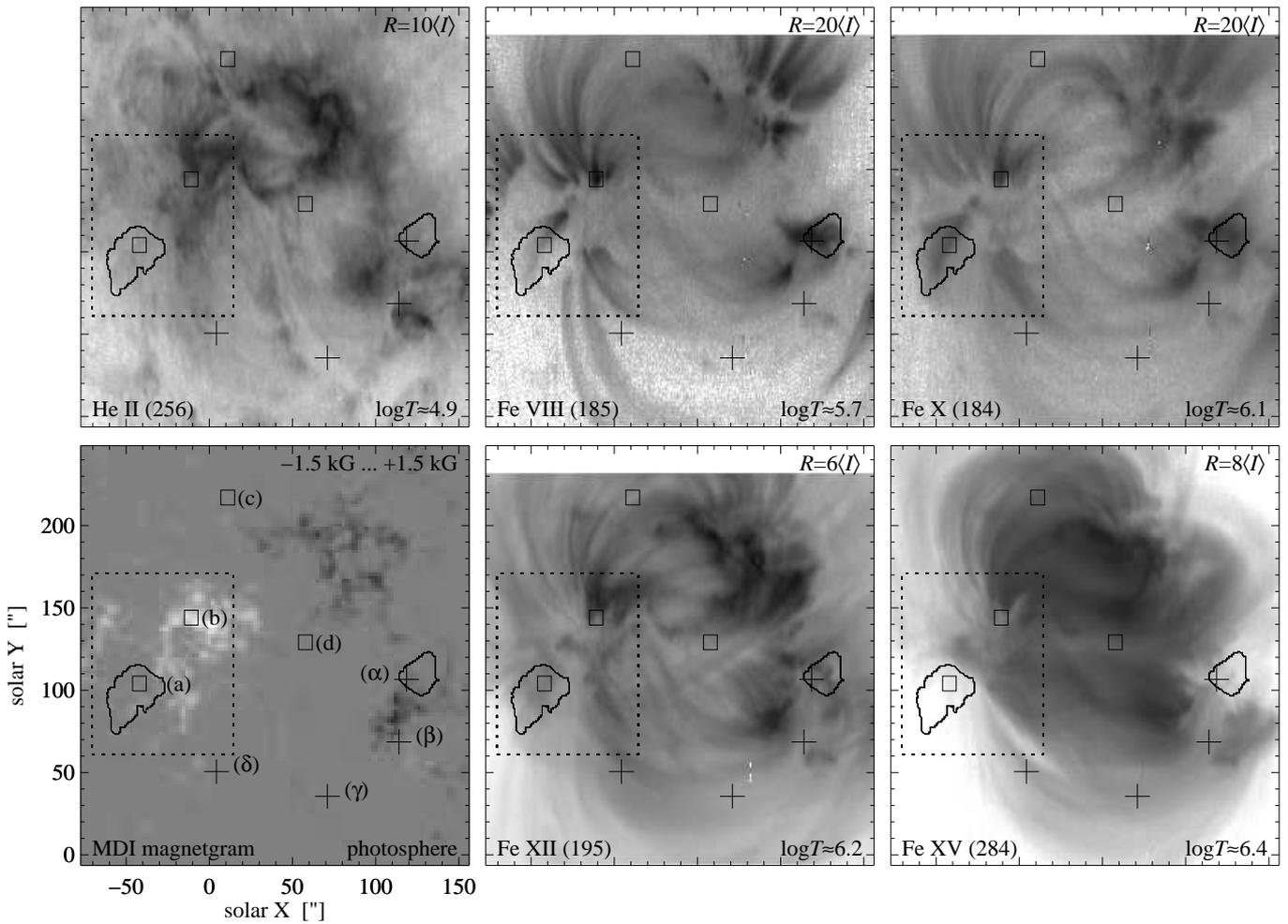}
\caption{%
Same as \fig{F:context}, but now the intensities are plotted on a logarithmic scale. The plotting ranges $R$ are different from \fig{F:context}.
\label{F:context.log}
}
\end{figure*}
%---------------------------------------------------------------------------

\end{document}